\documentclass{article}%
\usepackage{amssymb}
\usepackage{amsfonts}
\usepackage{amsmath}
\usepackage{graphicx}%
\providecommand{\U}[1]{\protect\rule{.1in}{.1in}}

\begin{document}

\title{Unexplored regions in QFT and the conceptual foundations of gauge theories }
\author{Bert Schroer\\CBPF, Rua Dr. Xavier Sigaud 150 \\22290-180 Rio de Janeiro, Brazil\\and Institut fuer Theoretische Physik der FU Berlin, Germany}
\date{June 2011}
\maketitle
\tableofcontents

\begin{abstract}
Massive quantum matter of prescribed spin permits infinitely many
possibilities of covariantization in terms of spinorial (undotted/dotted)
pointlike fields, whereas massless finite helicity representations lead to
large gap in this spinorial spectrum which for s=1 excludes vectorpotentials.
Since the nonexistence of such pointlike generators is the result of a deep
structural clash between modular localization and the Hilbert space setting of
QT, there are two ways out: gauge theory which sacrifies the Hilbert space and
keeps the pointlike formalism and the use of stringlike potentials which
allows to preserve the Hilbert space. The latter setting contains also
string-localized charge-carrying operators whereas the gauge theoretic
formulation is limited to point-like generated observables.

This description also gives a much better insight into the Higgs mechanism
which leads to a revival of the more physical "Schwinger-Higgs" screening idea.

The new formalism is not limited to m=0, s=1, it leads to renormalizable
interactions in the sense of power-counting for all s in massless representations.

The existence of stringlike vectorpotentials is preempted by the Aharonov-Bohm
effect in QFT; it is well-known that the use of pointlike vectorpotentials in
Stokes theorem would with lead to wrong results. Their use in Maxwell's
equations is known to lead to zero Maxwell charge. The role of
string-localization in the problem behind the observed invisibility and
confinement of gluons and quarks leads to new questions and problems.


\end{abstract}

\section{Introductory remarks}

Particle theory of the past century has led to a vast body of knowledge, but
many theoretical problems about the conceptual foundations of these
discoveries remained unresolved. Ideas as quark/gluon confinement have not
been understood in terms of quantum field theoretical interactions. Even less
ambitious looking problems as why and how the description of electrically
charged particles in quantum electrodynamics (QED) requires a noncompact
localization, whereas the Schwinger-Higgs screening counterpart of scalar
quantum electrodynamics is pointlike-generated, still lacks good understanding.

To get a feeling for the dimension of the problem in relation to completely
solved older foundational problems, compare this situation to the post
Faraday-Maxwell but pre Einstein era of classical electrodynamics. Already 20
years after its discovery, \textit{everything}, except the problem posed by
the ether, was in place; and when Einstein removed the ether from its throne,
no equation in Maxwell's theory and not even the Lorentz transformation had to
be modified. Although the removal of the idea of an ether by Einstein's theory
of relativity was an epoc-making event without which the emergence of quantum
field theory (QFT) is not imaginable, classical field theory and in particular
Maxwell's equation did not change their mathematical form. The action at the
neighborhood (Nahewirkung) principle was fully established; only its
connection with space and time was still awaiting drastic conceptual modifications.

The situation in QFT is very different; none of its fundamental equations for
interacting fields and particles has been brought under mathematical control.
In particular its best result, the renormalized perturbation theory, is known
to diverge. Even after almost 9 decades there is simply no conceptually closed
part of QFT which can be compared with the closure of Maxwell's classica ED.
The excellent conceptual understanding of quantum mechanics (QM) reached after
less than 3 decades shows that an explanation has to go beyond the shared
letter $\hslash$ and operators in Hilbert space.

When people say that QFT, in particular the standard model (SM), explain an
unprecedented amount of data, they refer to the power of prediction (sometimes
post-diction) which results from combining calculations based on perturbative
renormalization theory together with phenomenologically motivated assumption.
This is indeed impressive, but it should not create the illusion that QED, not
to mention the standard model (SM) into which it has been incorporated, has
reached the conceptual maturity after more than half a century which Maxwell's
theory attained already after only two decades. To say that quantum theories
are inherently conceptually more opaque does not help and is not convincing
since quantum mechanics (QM) in its epistemological and conceptual
understanding does not lag much behind classical field theory. In fact as a
result of its often counterintuitive structure, there is hardly any other area
of physics which had received as much successful attention as QM. Rather we
should admit that, although we have developed a detailed vocabulary within a
an impressive set of results obtained by a variety of computational tools, the
main impact has been to impress ourselves; from a comprehensive understanding
of the physical concepts of e.g. QED we are presently further away than the
discoverers of renormalization theory considered themselves to be. Being aware
about this state of affairs even with respect to our oldest and best studied
QFT, it is less disquieting that the central problems behind what we were
rather quick to call \textit{gluon/quark confinement}, and which thereby
acquired a kind of mental reality, persevered for more than 5 decades without
any essential progress.

In view of the impressive progress (renormalization theory including
Yang-Mills theories, dispersion relations, beginnings use of low-dimensional
QFT as theoretical laboratories) which was sustained well into the 70ies, this
raises the question about the reasons for the more than 4 decades lasting
stagnation. Are we less intelligent, have the problems become too difficult,
or has something gone wrong at an important conceptual crossing which forced
particle theory into a dead end? This will not be the subject of this paper,
we refer the reader to \cite{Ei-Jor}\cite{Causal}.

The present work is an attempt to bring some movement into a subject which,
although outside the range of the above critical remarks, is still far removed
from its closure. The only attempt with a similar aim was made by Mandelstam
when he tried to formulate the QED interaction in terms of field strengths
instead of point-like vectorpotentials. Our starting point is the recognition
that massless higher spin $s\geq1$ representations exhibit a certain
\textit{clash between localization and the Hilbert space setting} of quantum
theory. More specifically it is not a clash between the $(m=0,s\geq1)$
representations and unitarty as such, but an incompatibility of Hilbert space
positivity with the existence of pointlike generators with a prescribed
covariance behavior. In such a situation there are two ways out: to cede on
the side of Hilbert space or on the side of pointlike localization.

The first path leads to the gauge-theoretic formulation and is inexorably
related to indefinite metric spaces, whereas the second setting deals with
string-localized covariant potentials which interact with matter fields and
transfer their noncompact localization to the latter. The stringlike
localization of matter fields (Maxwell charges, Yang-Mills charges) belong to
what one could call the \textit{non-local gauge-invariants}. But this is only
a word since the ghost formulation of gauge theory is not capable to construct
such objects; it is simply not made for addressing any physical object unless
it is pointlike generated.

The gauge theoretic formalism has the advantage that its perturbative
formulation makes intimate contact with the formalisms of constraint classical
field theory as the BRST setting or the Batalin-Vilkovisky formalism which
permit to represent the constraint solution space in cohomological terms.
Although the protagonists of these cohomological formalisms for handling
constraints are quantum physicists, the idea to solve problems by first
enlarging the number of fields (by ghosts and other unobservable objects) and
then playing on cohomological properties in order to take care of constraints
is entirely classical\footnote{Classical covariant pointlike potentials are
perfect objects of a classicaal constraint formalism. The Hilbert space
requirement and the resulting delocalization is a pure quantum effect and only
happens for "potentials" in (m=0,s$\geq1$) representations.} and draws its
importance from the fact that the quantization approach (as an (artistic)
parallelism between classical and QFT) requires a good peparation on the
classical side. As the name "ghost" indicates, the enlarging objects which are
natural in their classical context lead to a breakdown of the Hilbert space in
attempts to quantize this formalism; but at least perturbatively the
compatibility with a Hilbert space representation \textit{after the
cohomological descent} can be established in terms of successful recipe of
"quantum gauge invariance" \cite{Scharf}.

There is however a high prize to be payed on the side of QT. Whereas the
intermediate abandonment of the Hilbert space can be accepted as a provisional
perturbatively convenient tool, the problem that the physically most important
electrically charged operators cannot be constructed in this perturbative
setting (because they are string-like generated, and classical Lagrangians
theory does not go beyond pointlike fields) is a more serious obstacle and
poses a challenge to look for something else.

This consists in the use of non-pointlike free fields which have no
Lagrangians \footnote{There are even many spinorial free fields which are not
of the "Euler-Lagrangian" kind. The latter are only needed in functional
integral inspired constructions.}, a step which places the localization
property on which QFT is founded back into the center stage. Localization is
important in QM and QFT and this carries the danger to overlook the enormous
conceptual-mathematical difference in their localization structures namely
that of the (Newton-Wigner\footnote{The addition of these names serves to
remind the reader that the suitably adapted localization connected with
probabilities also exists in the relativistic setting. In relativistic QM it
is the only localization \cite{interface1}.}) Born localization in QM and the
"modular localization" (quantum causal localization) for QFT. Elsewhere it was
shown how the confusion between the two has led particle physics into its
still ongoing crisis \cite{Causal}. The topic of this paper is more upbeat:
the modular localization will be used to shed new light on remaining problems
of gauge theory whose solution have a good chance to take the unfinished SM
out of its 40 year conceptual stagnation.

More concrete: it will be shown that modular localization resolves the clash
between pointlike localization and Hilbert space structure which affects
vector/tensor potentials of zero mass $(m=0,s\geq0)$ Wigner representations
and the resulting covariant semiinfinite string-localized potentials in a
QED-like interactions delocalize charged particles without affecting the
pointlike localization of field strength. This paper falls short of a
systematic perturbation theory; partially because important conceptual steps,
as the generalization of the Epstein-Glaser ideration, have yet to be accomplished.

Since the historical roots are not that well known anymore among novices of
QFT, the next section will review some of the old problems however with modern
hindsight. Often problems, which at the time were too difficult to be solved
in the realistic context, have been analyzed by looking at their analog in the
"theoretical laboratory" of two-dimensional QFTs. For the problem at hand this
will be commented on in section 3. The following section 4 contains the
representation theoretical setting. Together with modular localization THIS
leads to the interaction-free string-localized (s-l) covariant potentials.
Section 5 shows that already in the absence of interactions the use of
pointlike indefinite metric vectorpotentials is a risky business since in
contrast to the Hilbert space compatible s-l potentials they may lead to
incorrect results of which the absence of the well established quantum
Aharonov-Bohm effect (the violation of Haag duality for toroidal regions) is
an illustration.

In section 6 the main issue is to give perturbative arguments about the
transfer of delocalization from potentials to charged fields which thereby
loose their pretended pointlike nature. This section also presents the Higgs
mechanism in a different more physical setting of Schwinger-Higgs screening
which leads to the re-localization of the nonlocal aspects of Maxwell charges
in analogy to Debeye's screening in QM which converts the Coulomb potential
into an effective short range interaction of the Yukawa kind. The new aspect
is that screening in QFT is not just a vanishing of the global charge operator
on physical states, but also involves the return from s-l to point
localization. In the last section some more speculatutive points of view about
confinement/invisibility of states in connection with nonabelian gauge
theories are mentioned before some of the high points are resumed in the
concluding remarks.

\section{History of electrically charged fields and the problem of
infraparticles}

The gradual improvement of the understanding of the local quantum physical
aspects of electrically charged fields and their associated particles is one
of the most fascinating projects of QFT. Even after a resounding success in
describing observed data dating back more than 70 years, is not anywhere near
its conceptual closure. It leads to a particle-field relation which is far
more subtle than the standard textbook case of an energy-momentum spectrum
with mass gaps. In the latter case one obtains in a well-known manner (via
large time asymptotes) free fields; together with the concomitant assumption
of asymptotic completeness\footnote{All presently known constructions are
based on the two-dimensional "bootstrap-formfactor setting" (factorizing
models). For those constructions which start from particles and introduce
fields through their formfactors \cite{Ba-Ka}, the completeness property is
built into the construction.} one then finds that the Hilbert space of such a
theory has the form of a Wigner-Fock space.

The appearance of zero mass particles as such do not necessarily cause a
breakdown of these scattering results. As long as the particle states in the
presence of interactions manage to preserve their particle aspect i.e. are
\textit{not} "sucked" into a continuous part of the mass spectrum (example the
particle properties of the nucleon are not affected by the interaction with
zero mass pions) the results of scattering theory continue to be valid, even
though there are no mass gaps.

But interactions of charged particles with photons are not covered by this
kind of scattering theory. Already the quantum mechanical Coulomb scattering
leads to problems with the large time behavior of amplitudes, although in this
case the multiparticle tensor structure of the Hilbert space remains
unaffected and the large time limits converge after removing a logarithmic
time dependent phase factor \cite{Dollard}. In QED the failure of
time-dependent scattering theory is more severe, in fact it is inexorably
related to a breakdown of Wigner particle states and the absence of a
Wigner-Fock structure of the QED Hilbert space i.e. it is not limited to
multi-particle scattering amplitudes but even changes the nature of
one-particle states themselves. Instead of the mass shell contribution in the
K\"{a}llen Lehmann twopoint-function one finds a branch cut whose strength is
restricted by unitarity and which leads to a vanishing LSZ limit making it
impossible to define a nontrivial scattering matrix. On the other hand in
perturbation theory the restriction to the mass shell causes infrared
divergencies which, unlike the Coulomb scattering in QM. \textit{cannot be
repaired} on the level of scattering amplitudes but only by passing to
infinite soft photon-inclusive cross sections with infrared cutoffs in
intermediate steps and a dependence on the method by which the cutoff is
introduced and afterwards compensated. In contradistinction to the elegant
spacetime representation of the scattering amplitudes in the LSZ formalism
there is (yet?) no known spacetime representation.

It is the principle aim of this paper to establish that these unusual momentum
space properties are caused by the string-localization of vector potentials
which, figuratively speaking, feed their string localization through the
interaction to the charged matter fields. Whereas for the potentials
themselves the effect of string localization has no direct physical
consequence as long as there is no direct coupling between them (abelian gauge
theories) and the relation to the pointlike field strength remains unaffected,
the charged particles, which were described by pointlike matter fields in zero
order perturbation order, become semiinfinite string-localized in a way which
cannot be undone in any operational way (differentiation or other lin.
operations). This weakening of localization of quantum Maxwell charges is
behind the \textit{infraparticle} concept. But before we get there we will
follow briefly the historical path.

Problems with the application of standard scattering theory to QED were
noticed quite early since scattering theory in one form or other is one of the
oldest tools in QT\footnote{The Born probability concept was first introduced
in the setting of Born's scattering approximation, the extension to
Schroedinger wave functions came later.}; the research on infrared
divergencies begun in a 1934 paper by Bloch and Nordsiek in which scattering
of charged particles and photons was analyzed in a simplified model of QED
\cite{B-N}. The important conclusions, namely that although the number of
photons in the infrared is infinite, the emitted energy and angular momentum
remains finite, as well as the message that, in agreement with the vanishing
of the LSZ scattering amplitudes, one has to sum over all infinitely many
infrared photons up to a certain energy resolution $\Delta$ determined by the
measuring device in order to obtain a non-vanishing scattering probability
(inclusive cross section\footnote{In agreement with the vanishing of the LSZ
amplitudes, one obtains vanishing of the probabilities in the limit
$\Delta\rightarrow0.$}), can already be found in this early paper. Later
perturbative covariant calculations in QED succeeded to confirm and improve
these results \cite{YFS}.

If QFT would be limited to the finding of successful recipes, then this
formalism, which proceeds "as if" charged particles would be Wigner particles
and the LSZ asymptotic behavior would be valid modulo some infrared
imperfections which can be repaired similar to the removal of exponential
logarithmically diverging long distance factors in Coulomb scattering. But the
infrared trouble is not limited to scattering amplitudes, its course is the
breakdown of \ Wigner's one-particle structure and the possibility to describe
QED in a Wigner-Fock space. Electrically charged quantum matter is
semiinfinite string-localized in a very sense, the infinite string has wiped
out the one particle pole $p^{2}=m^{2}$ and replaced it by a more complicated
mass spread which has no characterization in terms of just two invariants
$m~and~s$. For many pragmatic minded particle theorists the compensation of
infrared divergencies on the level of soft photon inclusive cross sections
would have been the end of the story. Fortunately theoretical physics was
never pragmatic in this extreme sense, since one knows that suppressed
conceptual and philosophical questions always return later on with a
vengeance. \ 

In this paper it will be shown that the infrared divergences are the
\textit{result of the presence of string-localized vectorpotentials} in the
zero order interaction density. Their most important role is to
\textit{delocalize charge fields} with which they interact and as a result to
lead to the phenomenon of the charged "infraparticles" which results in a
continuous mass spread accumulated at the lower edge of a continuum. The
effect on the string-localized vectorpotentials on themselves on the other
hand remains very mild, even in the presence of interaction the associated
field strength stay pointlike and the scattering theory for photons only
remains close to the conventional setting \cite{Haag}. This behavior occurs
for all zero mass, $s\geq1$ potentials but not for $s\leq1/2$ which includes
the aforementioned zero mass meson to nucleon coupling. The the reason behind
the appearance of infraparticles is the semiinfinite string localization of
the vectorpotentials which the interaction transfers to the charged particles
and the dissolution of the mass shell of the charged particle into the photon
continuum is the momentum space consequence. In the presence of
selfinteractions between vectorpotentials (e.g. Yang-Mills) there are no
linearly related field strengths and the rules for change of the string
localization becomes more complicated (interaction-dependent).

According to the previous remarks the two-point function of a charged
field\footnote{With charged field we always mean the physical charged field,
but it is not necessary to add this since the Lagrangian matter field $\psi$
in the indefinte metric setting has neither electric chage nor has its
localization any physical significance.} in interaction should reveal a
behavior at the mass shell which is different from a delta function. Unitarity
imposes a strong restriction on the two-point function. For example
derivatives of delta functions are excluded, since they are not positive
measures; a moment of thought suggests that near $p^{2}\approx m^{2}~$\ the
Kallen-Lehmann function should have an anomalous (coupling-dependent) power
and behave as $\theta(p^{2}-m^{2})\left(  p^{2}-m^{2}\right)  ^{f(e)}%
g(p^{2},\alpha)$ which for vanishing coupling (fine structure constant)
$\alpha\rightarrow0$ approaches the mass shell delta function and for finite
$e\neq0,$ for reasons of representing a positive measure, is milder than the
delta function. Physical QED (i.e. not its indefinite metric version) leads to
a spontaneous breaking of Lorentz invariance in its charged sectors, but since
in perturbation theory the calculation of correlations involving physically
charged fields is prohibitively difficult, the normal practitioner does not
see such things.

In the later part of the present work it will be shown that there is an
additional vector in the problem which in principle also enters the infrared
powers. This state of affairs (a milder mass shell singularity than a delta
function) would immediately lead to a vanishing LSZ limit, which accounts for
the observation by Bloch and Nordsiek about the vanishing of scattering
leading to a finite number of photons after the compensation of the infrared
cutoffs have taken place.

Since QED is hard to control, physicists began in the 60s to look at simple
soluble two-dimensional \textit{infraparticle models} which exhibited the
expected coupling-dependent power behavior, similar to that which appeared in
the scattering formulas of YFS \cite{YFS} after summing over leading
logarithms in the infrared photon cutoff. A common feature of those
two-dimensional infraparticle models is the presence of an exponential zero
mass two-dimensional Bose field factor in the generating field \cite{S1}. What
was still missing was a structural argument that electrically charged
particles are really infraparticles in this spectral sense, as well as a
conceptual basis for a new scattering theory which does away with cutoff
tricks and, as the standard LSZ or Haag-Ruelle scattering theory, only uses
spacetime properties of correlation functions.

In QFT the overriding principle is causal localization and, thinking of the
relation between energy positivity and localization (sections 3,4), it is in a
certain sense the only one. This principle has a variety of different physical
manifestations and the main problem of understanding a model of QFT consists
in \textit{finding the correct structural arguments which reveals the
connection between the perceived properties of a model and their explanation
in terms of causal localization}.

The role of localization can be exemplified in two physically very important
cases (section 5), the string-localization of electric charges and the
Schwinger-Higgs charge screening as a kind of "re-localization" process,
leading to a vanishing charge (a loss of the charge superselection) causing a
breaking of the charge symmetry. In the case of the before mentioned
two-dimensional infraparticle models this consists in the realization that the
complex zero mass Bose field is really a semiinfinite string-like localized
field (next section) and it is this noncompact localization behavior which is
at the root of the dissolution of the mass shell delta function into a\ cut
type singularity. Computations may be easier in momentum space, but the deeper
conceptual insight is always in the spacetime setting.

Suggestions that electrical charge-carrying fields as the interacting
electron-positron field have necessarily a noncompact extension entered the
discussion quite early; practically at the same time of the Bloch-Nordsiek
work Pascual Jordan \cite{Jordan} started to use the string-like formal
presentation of the physical charge field which, since this also was on
Dirac's mind and is also often linked with the later work by
Mandelstam\footnote{Mandelstam's \cite{Man} use of line integrals over field
strength is certainly an early attempt to preserve the Hilbert space structure
by easing on pointlike locality. But the central position of causal
localization in QFT was not yet fully recognized.}, will be referred to as the
DJM presentation of a charged field.%

\begin{align}
\Psi(x;e)  &  =~"\psi(x)e^{\int_{0}^{\infty}ie_{el}A^{\mu}(x+\lambda e)e_{\mu
}d\lambda}"\label{DJM}\\
\Phi(x,y;e)  &  =\ "\psi(x)e^{\int_{0}^{1}ie_{el}A^{\mu}(x+\lambda
(x-y))(x-y)_{\mu}d\lambda}\bar{\psi}(y)" \label{br}%
\end{align}
Gauge invariance not only suggested that physical\footnote{The pointlike
formal matter fields which enter the Lagrangian and field equation of gauge
theories are auxiliary quantities which act neither in a Hilbert space nor
does their localization have a physical significance.} electrically charged
fields have a noncompact localization with the semiinfinite spacelike string
(\ref{DJM}) being the tightest (least spread) possibility, but also that
charge-neutral pairs are necessarily interconnected by a "gauge bridge"
(\ref{br}). The methods of local quantum physics permit to show the infinite
extension of charges on the basis of a rigorously formulated quantum Gauss law
(next section), but the above formula has no conceptual or computational
preferential status and is not distinguished by renormalization theory; in
fact \textit{a physical (Maxwell) charged field does not appear at all within
the standard (Gupta-Bleuler, BRST perturbative formalism} \cite{F-Mo-Str}%
\cite{Mo-Str}. This zero Maxwell charge effect in the presence of ghosts is an
analog of the zero magnetic Aharonov-Bohm effect in such a setting (section
5). Formally the correct result would be obtained after imposing the BRST
invariance but nobody knows how to do this for nonlocal expressions; the
classical BRST formalism is only made to construct local observables. In a
Hilbert space description with string-like potentials these effects are
correctly described and the problem is shifted to the renormalization theory
in the presence of string-localized free fields.

The rigorous definition of these charged fields in renormalized perturbation
theory is a nontrivial problem (the reason for the quotation marks in
(\ref{DJM})). Steinmann had to develop a separate perturbative formalism only
for defining the renormalized physical DJM charged fields \cite{Stei}. The
gauge setting leads to a nice picture suggesting the semiinfinite
string-localization of charged fields, but, just as the abstract argument
based on the quantum Gauss law, it does not really explain the origin of this
weaker localization in terms of the form of the interaction. In the standard
gauge setting the latter is pointlike i.e. it looks like any other local
interaction. This shifts the problem of physical localization to a level where
definitions are cheap but their constructive use turns out to be difficult.
The perturbation theory of QED in the BRST setting follows standard rules, but
the calculation of BRST invariant correlations is prohibitively difficult if
nonlocal operators enter the search; even applied to local operators it is
difficult since the BRST transformations resembles that of a nonlinear acting
symmetry. In fact most of the papers present the formalism, but fall short of
calculation BRST invariant correlations of charged objects.

Needless to add that for Yang-Mills interactions even the first step is beset
by inexorably intertwined ultraviolet-infrared divergencies which impede the
execution of the renormalization program in any covariant gauge. \ Ultraviolet
divergencies without intermingled infrared problems can be renormalized which
expresses the fact that in a more intrinsic approach which avoids the quantum
mechanical aspects of the standard (cutoff, regularization) methods in terms
of a more intrinsic (Epstein-Glaser) approach (which takes better care of the
intrinsically singular structure of quantum fields resulting from vacuum
polarization). A particular elegant formulation is the "algebraic adiabatic
limit" method \cite{Fred} which is based on the dissociation of the algebraic
structure from that of states. The latter step is the analog of the DHR
superselection construction but the existing theory \cite{Haag} is not
applicable to Maxwell charges. Infrared divergencies which hide physical
localization problems exist for all theories in which massless higher spin
$s\geq1$ potentials participate in the interaction, independent of the
presence of ultraviolet problems. Their control is not a matter of some
mathematical-technical adjustments but rather a major revision of the physical
setting. The string-like formalism leading to a distributional directional
dependence allows to seperate them.

The fact that the fundamental electrically charged fields cannot have a better
localization as semiinfinite stringlike, has of course (at least implicitly)
been known since the DJM formula (\ref{DJM}), but the infrared problem showing
up in scattering and the problem of semiinfinite string-localized charged
fields have for a long time not been linked together. It is one of the aims of
this paper to show that they represent two sides of the same coin.

The standard gauge formalism is, apart from some conceptual
problems\footnote{Different from non-gauge QFT the Hilbert space for the local
observables is not defined at the outset but rather results from an auxiliary
indefinite Hilbert space through a cohomological construction.}, very
efficient in setting up a perturbative formalism for \textit{local} gauge
invariants. As mentioned one of the reasons why this formalism is not suitable
to formulate the problems behind the infrared divergencies is that it does not
give any clue how to deal with electrically charged operators; they are just
not part of any existing perturbative formalism. Attempts to attribute
physical significance to the Dirac spinors in the indefinite metric formalism
have ended in failure; there is simply no subspace on which the Maxwell
equations can be defined and on which states with a nontrivial electric charge
can be introduced, the pointlike $\psi^{\prime}s$ carry no Maxwell charge and
their pointlike localization is a fake \cite{F-Mo-Str}\cite{Mo-Str}.

If the underlying philosophy of local quantum physics (LQP), which led to the
construction of charge superselection sectors solely from the structural data
of observables, would also apply to Maxwellian charges, one should be able to
reconstruct the charge neutral bilocals with gauge bridges (\ref{br}) from the
algebra of charge-neutral pointlike local bilinears. For globally charged
fields (in which case there are no connecting gauge lines) this has been shown
\cite{L-S} in a model, but for Maxwellian charges there are yet no
mathematically decisive result \cite{Jac}. But gauge bridges are much more
removed from what can be reached on the collocational radar screen of standard
gauge theory than string-like potentials.

We propose a new approach for theories in which string-localized potentials
associated to zero mass finite spin representations interact with massive
quantum matter. This includes in particular models of gauge theories. The new
setting is designed to incorporate the physical charged fields into the
perturbative formalism. Figuratively speaking, the potentials which have a
rather harmless string dependence pass the string-localization onto the
massive field where it keeps piling up in perturbation theory and (in contrast
to the potentials themselves) becomes \textit{irremovable by any linear
operation} already in the lowest nontrivial order (section 5).

The starting point is a combination of Wigner's representation theory for
$(m=0,s\geq1)$ with modular localization. The result is that, whereas the
various possible "field strengths" are pointlike covariant wave functions (or
pointlike quantum fields in the functorial associated "second quantization"),
their "potentials" are semiinfinite stringlike localized objects. The
stringlike potential setting solves also another problem. For $s\geq1$ the use
of field strength does not allow interactions which are renormalizable in the
sense of power-counting, but this interdiction does not hold for couplings
with stringlike potentials; there always exist polynomial interactions of
maximal degree 4 in terms of stringlike potentials of short distance scaling
dimension $d_{sc}=1$ which are renormalizable by power-counting. Whether
desired interactions (as the Einstein-Hilbert action for s=2) are among those
is a separate question.

Textbooks and review articles on gauge theories in general (and on QED in
particular) often create the impression that they represent the best
understood QFTs. It is certainly true that they are the physically most
important models and they lead to rich perturbative calculations. It is also
true that in the semiclassical form of quantum theory in external gauge fields
they led to deep mathematics and broadened the level of mathematical knowledge
of several generations of theoretical physicists, but those geometrical
structure (fibre bundles, cohomology), contrary to a widespread
opinion\footnote{They did however raise the level of mathematical
sophistication of particle physicists.}, are not the kind of structures which
are important for the solution of the above quantum problems. Nevertheless
both the (semi)classical and the local quantum physical aspects of these
models are both very rich in their own right. They are certainly the most
interesting theories, especially in the setting which we are going to present.

As mentioned, the models which are conceptional well-understood are those
which have a\ mass gap and hence fall into the range of applicability of the
LSZ/Haag-Ruelle scattering theory; but after the acceptance of the standard
model those models have been of a lesser observational and conceptional interest.

There are also theories which, at least at the outset, have a somewhat hidden
mass gap than those which served as illustrations of the LSZ formalism. These
are the models which owe their mass to the Schwinger-Higgs screening mechanism
and play an important role in the standard model. They involve massive
vectormesons, but more generally all models which rely on interactions with
higher ($\geq1$) spin objects are interesting because the pose a challenge to
renormalization theory, as it will be shown in section 3.

Schwinger's contributions to the screening idea has been forgotten, therefore
some historical remarks are in order. At the end of the 60s Schwinger
envisaged the possibility of a "screened phase" in actual (spinor) QED in
which the photon becomes a massive vectormeson. Since I could not find any
convincing argument, he looked at d=1+1 massless QED (the Schwinger model)
where he could exemplify his screening idea \cite{Schwinger}; the model led to
a vanishing (totally screened) charge showing also that the screening
mechanism is not related to the Goldstone spontaneous symmetry breaking since
the latter has no realization in d=1+1. The screened version of scalar QED in
d=1+3 is identical to the model proposed by Higgs \cite{Higgs}. A screening
mechanism for s=1/2 quantum matter in d=1+3 does not exist, at least not in
perturbation theory. Therefore one always needs to start with scalar QED or
with couplings of Yang-Mills fields to complex scalar fields in order to
obtain massive vectormesons interacting with Schwinger-Higgs screened (real)
scalars. If other fields as s=1/2 Dirac fields are coupled to massive
vectormesons obtained from screening, the only charge which survives is the
global charge carried by Dirac spinors whereas the Maxwell charge and its
nonabelian counterpart vanishes.

We will limit our main attention to two models namely scalar QED with its
delocalized (string-localized) charged fields, and associated infraparticles,
and the Higgs model, which in some sense to be made precise may be viewed as
the result of "charge screening" leading to mass gaps and a return to
pointlike locality. With the exception of this \ "mass-generating"
Schwinger-Higgs screening mechanism, all gauge theories contain strongly
delocalized objects, namely visible Maxwell type charges or invisible gluons/quarks.

In the following it will be shown that the infrared divergence properties and
their first cure in the famous Bloch-Nordsiek \cite{B-N} scattering model and
Jordan's stringlike (\ref{DJM}) formula which represents a physical
electrically charged field, are really two different sides of the same coin;
they are both consequences of the fact that certain quantum objects by their
intrinsic nature do not permit pointlike generators but rather possess only
noncompact localized generators. As the family of arbitrarily small double
cone\footnote{A spacetime double cone is the unique kind of spacetime region
which is compact, simply connected and causally complete.} localization (the
natural shape of a compact causally closed region) is pointlike generated, and
the tightest causally complete noncompact localization, namely a thin a
spacelike cone, has as its core a semiinfinite spacelike semiinfinite string.

It may be more than a curiosity that both observation, the one on infrared
divergencies in scattering of charged particles, and the stringlike DJM
formula were made at practically the same time. Couldn't it be that there is
an aspect of the subconscious in the Zeitgeist? Hard to decide because Jordan
used the DJM formula mainly for deriving an algebraic magnetic monopole
quantization in the same year that Dirac presented his geometric derivation.
Historical speculations aside, the understanding of why these two observations
belong together is more recent and constitutes a strong motivation for the
present work.

The content is organized as follows. In the next section we review the
two-dimensional infraparticle models and show that all of them are
string-localized (half-space-localized). The section also contains the known
rigorous statements about higher dimensional infraparticles.

Section 3 presents the theory of string-localized potential fields starting
from Wigner's representation theory and illustrated in more detail for s=1, 2.

The last section presents some rudiments of a new formalism, in which instead
of enforcing pointlike potentials and paying the prize of being temporarily
thrown outside quantum physics (by the occurrence of indefinite metric), one
rather sticks to a setting in which one confronts the true stringlike
localization instead of working with a simpler and familiar pointlike
formalism which, apart from local BRST invariants does not describe the true
localization of some of the most important physical objects a charged fields.

\section{Lessons from 2-dim. infraparticles}

The Bloch-Nordsiek treatment of the infrared aspects of scattering of
electrically charged particles and its extension to full perturbative QED in
the work of Yennie, Frautschi and Suura \cite{YFS} led to the notion of
infrared finite \textit{inclusive cross sections} in which infinitely many
"soft" photons are summed over. This successful recipe did however not answer
certain important conceptual questions. The quantum mechanical treatment of
Coulomb scattering also leads to an infrared problem, namely a scattering
theory of charged particles in which for large times a logarithmic
time-dependent phase factor prevents the asymptotic convergence for large
times \cite{Dollard} and where the remedy is either the removal of this factor
from the amplitudes or passing directly from amplitudes to probabilities. In
this case one knows in spacetime dependent terms what one is doing, however a
manipulation in momentum space as in YFS is a recipe and only gains a
conceptual status one finds a spacetime explanation for what one is doing.

So the question arose: is the QFT scattering of charged particles similar to
the quantum mechanical Coulomb scattering in which one particle states and
their n-particle tensor-products continue to exist in the Hilbert space of the
theory and only the large time asymptotic convergence towards n-particle
states is modified by infrared factors, or is there something more
dramatically happening in the QFT scattering of charged particles?

The field theoretic phenomenon of vacuum polarization in the presence of
interactions leads to the mutual coupling of all channels as long as they are
not separated by superselection rules. Since relativistic scattering theory
treats one-particle states and multiparticle states on the same footing, one
would expect that, different from the Coulomb scattering, radical changes in
the scattering concepts can not happen without a major modification of the
particle concept, so that even a charged one-infraparticle state cannot be
described as a kinematical object in terms of Wigner's irreducible
representation theory. Since, as most QFTs in d=3+1 which have no spectral
gaps, QED is still outside mathematical-conceptual control, it has been useful
to look for analogs in the more accessible two-dimensional "theoretical
laboratory". QED shares with all other renormalizable theories in any
dimension the divergence of the perturbative series so that great care has to
be applied in drawing structural conclusions.

The exponentiation of leading logs to a power behavior in the YFS work
suggested to consider models which contain a zero mass exponential Bose field
$\Phi\footnote{Interestingly enough, this was also a model used by Jordan
though not for the present purpose but rather for developing his ill-fated
"Neutrino theory of light".},$ which in d=1+1 has formal dimension zero and
therefore leads (still formally) to logarithmic correlation functions of
exponential operators with an anomalous operator dimension
\begin{align}
&  \dim e^{i\alpha\Phi(x)}\simeq\alpha^{2}\\
\mathcal{L}_{int}  &  =\alpha\partial_{\mu}\Phi\bar{\psi}\gamma^{\mu}%
\psi,\ \psi=\psi_{0}(x):e^{i\alpha\Phi(x)}: \label{int}%
\end{align}
Historically the first use of a Lagrangian model as an "theoretical laboratory
for the infrared" \cite{S1} was a zero mass scalar meson coupled to a massive
nucleon via a derivative coupling (\ref{int}) but actually this exponential
should be called the Jordan model \cite{Jordan} since Jordan invented it at
the same time of the Bloch-Nordsiek papers and his line integral presentation
of gauge invariant charged fields (\ref{DJM}) though for a very different
purpose\footnote{At that time physicist thought that, as in QM, one can study
particle phenomena in low dimensions and the simply generalize to higher ones.
But from Wigner's work on particles we know that the representation theory of
the Poincare group is very much dimension dependent and hence to infer from
the chiral Bosonization-Fermionization relation a "neutrino theory of light"
is really far-fetched \cite{Jordan}.}. The resulting "interacting" Dirac
spinor (\ref{int}) leads to a Kallen-Lehmann representation in which, instead
of the one-particle mass shell delta function, one encounters a cut which is
starting at the position of the mass and for $\alpha\rightarrow0$ converges in
the sense of distributions to the mass shell delta function. The strength of
this cut (the power) is bounded by unitarity since the latter forces the K-L
weight to be a measure. The free Dirac field $\psi_{0}$ is a formal auxiliary
object; in the autonomous Hilbert space constructed via the GNS reconstruction
from the Wightman functions of $\psi,\bar{\psi}$ spinor fields there is no
free $\psi_{0}$ field and the unitary representation of the Poincar\'{e} group
does not contain an irreducible component corresponding to a discrete mass.

Consistent with this structure is the observation that the LSZ large time
asymptote vanishes i.e. instead of the standard incoming/outgoing free fields
one obtains zero. This is so because the infraparticle singularity is too weak
in order to match the dissipative behavior of particle wave function; in
perturbation theory one encounters however the typical well-known logarithmic
infrared divergencies. The non-perturbative model of Bloch and Nordsiek, as
well a the summing up of leading terms in the YFS work, lead to vanishing
emission amplitudes for the emission of a finite number of photons, which, as
mentioned previously, is in agreement with the vanishing of the LSZ limit as a
result of the softening of the mass-shell singularity mentioned before.

The field $\Phi$ (\ref{int}) has infrared properties which prevent it from
being a Wightman field, since it cannot be smeared with all Schwartz test
functions but only with those whose total integral vanishes. However the
exponential is again a bona fide Wightman field if one imposes on the Wick
contraction rules the charge superselection rule i.e. if one assigns to the
exponential the charge $\alpha$ and to its adjoint the charge -$\alpha,$ so
that the only products of fields with total charge zero have nonvanishing
vacuum expectations.

The selection rules would follow from the exponential of a massive spinless
field in the massless limit by imposing the condition that the fields are
renormalized with appropriate powers of the mass in such a way that none of
the correlations becomes infinite. This requirement is well-known to lead to
the vanishing of all correlations of exponential fields except those involving
charge neutral products. In the above derivative model (\ref{int}) this local
$\alpha$-selection rule is masked due to the presence of a second
\textit{global} charge conservation of a complex Dirac spinor.

For our purposes another method, which shows that the exponential field is
\textit{string-localized,} leads to more physical insight. It starts from a
chiral current which is a well-defined quantum field and defines the
exponential field as an exponentiated integral of the current over a
\textit{finite} interval followed by the spacetime limit which takes one of
the endpoints to $+\infty$ infinity in order to lessen the influence the
compensating charge has on the physics of isolated localized charges. The
correlation function only remain finite if the all the compensating charges at
infinity and hence all the finitely localized charges add up to zero. This
model indicates that charges, that there are certain charges which owe their
existence to string localization and the claim is that this simple exponential
Boson model illustrates certain features which in the realistic context of QED
are much harder to demonstrate.

This method reveals that the resulting operators are localized along a string,
namely the semiinfinite interval $\left[  x,\infty\right]  $ where x is the
endpoint which has been kept fixed. So formally the $\Phi$ in the exponential
should be viewed as%
\begin{align}
\Psi_{\alpha}(x)  &  =e^{i\alpha\int_{x}^{\infty}j(y)dy}\label{s}\\
\Psi_{\alpha}(x,x^{\prime})  &  =e^{i\alpha\int_{x}^{x^{\prime}}j(y)dy}%
\end{align}
The main difference to (\ref{DJM}), apart form the spacetime dimension is that
the current is a physical massless field whereas the pointlike vectorpotential
is not. It is quite easy to show that all chiral correlation functions,
including the charge superselection rule, can be obtained from
string-connected bilocals as defined in the second line (\ref{s}). The
$\alpha$-charge has similarities with a Maxwell charge in that it is locally
generated. In spinor QED there is besides the Maxwell charge also the global
charge\footnote{"Global" in this context does not mean that there are no local
properties or consequences. In the DHR theory \cite{Haag} of superselected
charges and inner symmetries, the global spinor charge is reconstructed via
the (representation-theoretical) "shadow" it imprints on the charge-neutral
local observables.}. But, as will be seen later, a Schwinger-Higgs screening
where both the Maxwell charge also the notion of global charge as well as the
charge which comes with a complex field disappear is only possible with scalar
complex fields i.e. scalar QED. Needless to add, the zero chiral field plays a
special role; it does not really exist rather what is behind it is a
string-localized object which only makes sense in the exponential form.

String localized fields, as those exponentials, are outside the standard
Wightman field theory in that they neither commute nor
anticommute\footnote{For the particular value of the charge $\alpha=1/2$ (it
depends on the normalization of the current $j$) one obtais the massless d=1+1
free Weyl spinor. For this reason the moving between $\Phi$ and this spinor
$\psi$ has been called fermionization/bosonization, a terminology which is not
completely correct since both live in different charge sectors.} for spacelike
separations between endpoints $x$. In chiral conformal field theory as
(\ref{s}) the string localization is visible in \textit{plektonic} (in the
above abelien case \textit{anyonic}) commutation relations representing braid
group statistics. Just looking at the problem as a formal massless limit of a
two-dimensional exponential would not reveal that behind the infrared
divergences there is a transition from point- to string-like localization. The
global Dirac charge and the local charge carried by the exponential
line-integral in the above derivative model (\ref{int}) lead to identical
selection rules and have their counterpart in QED where the local charge is
referred to as the Maxwell charge. In higher dimensions the string-like
localized nature of charged fields is more easily perceived, since fields
$\Psi(x,e)$ where $e$ is the direction (spacelike unit vector) of the
semiinfinite spacelike string $x+\mathbb{R}_{+}e$ have a Lorentz
transformation law in which the $e$ participates, whereas in d=1+1 there is no
such variable direction. As usual for spontaneouly broken symmetries, the
Lorentz transformation on d=1+3 charged fields exists only as an algebraic
automorphism (which also acts on the string direction) which cannot be
globally unitarily implemented.

The two-dimensional infraparticle models of the 60s and 70s have been recently
re-discovered in order to illustrate a proposal in the setting of effective
QFT called "unparticles". It is unclear how those models can illustrate two
infrared concepts unless they are the same. Unfortunately the authors have
only sketched in vague perturbative momentum space setting what they require
of unparticles. Apparently they believe that the Bloch-Nordsiek and YFS work
only affects the scattering amplitudes but leaves particle states intact
similar to Coulob scattering in QM. But this is not true, behind the infrared
problems in QED there are string-localized infraparticles and not just long
ranged interaction potentials leading large time exponential logarithmic
factors which are easily taken care of \cite{Dollard}. The unparticle proposal
is based on an insufficient appreciation of the radicality of infraparticles.
Writing down something in momentum space and naming it "effective QFT" without
giving a hint what one wants in terms of localization is not revealing much.

The concept of infraparticles is not explained by just pointing to momentum
space properties of Kallen-Lehmann spectral function; its conceptual pillar is
rather the weakening of localization of charge-carrying fields as a result of
de-localization through interactions with string-localized vectorpotentials
(see next section). The infrared anomalous power cut in the K-L spectral
function starting at the mass of the charged object as well as other momentum
space anomalies are consequences and not the cause of the unexpected deviation
from particle behavior (unexpected at least from the Lagrangian gauge theory
viewpoint which does not reveal any eye-catching particularity). Whereas in
d=1+1 formal zero mass scalars play a prominent role, in d=1+3 conformal
scalars with anomalous dimensions cannot generate the noncompact localization
which one needs to get the typical power cuts in the K-L two-point function;
one really need string-like potentials appearing in the interaction density
(next section)

Since most of the perturbative arguments in d=1+3 about Maxwell charges
inherit the mentioned problems of charged fields in the gauge theoretic
setting, it is useful to know that there exists a rigorous conceptual argument
based on a quantum field adaptation of the Gauss law \cite{Bu}. The arguments
can be found in Haag's book, it may however be helpful to remind the reader of
its main content. One starts from a $t-r$ smeared field strength%
\begin{align}
F^{\mu\nu}(f_{R})  &  =\int f_{R}(t,r)dtdr\int F^{\mu\nu}(t,r,\theta
,\varphi)f_{2}(\theta,\varphi)d\theta d\varphi\\
f_{R}(t,r)  &  =R^{-2}f(\frac{t}{R},\frac{r}{R})\nonumber
\end{align}
The test function smearing of the singular pointlike fields is necessary in
order to have a well-defined operator, for the classical field strength this
would be superfluous. The angular integral defines an operator which
represents the average flux through a sphere of radius r at time t. There are
two physically motivated assumptions about the charged state of interest
$\omega$%
\begin{align}
\omega(F^{\mu\nu}(f_{R}))  &  \neq0\label{Ha}\\
\omega(F^{\mu\nu}(f_{R})^{2})  &  <\infty\nonumber
\end{align}
which is interpreted as a consequence of a quantum adaptation of Gauss's law
i.e. the expectation value of the flux in a charged state deviates
significantly from the vacuum, but the fluctuation should remain as bounded as
they are in the vacuum since the correlation of the field strength for large
distances should not be influenced by the presence of a charge. Whereas the
previous formulae specify the mathematical definition of electromagnetic flux
through a spatial surface and charged state, the following commutation
relation between the mass operator and the averaged field strength is at the
heart of the matter:
\begin{equation}
\left[  M^{2},F^{\mu\nu}(f_{R})\right]  =iR^{-1}(P^{\sigma}F^{\mu\nu
}(f_{\sigma})_{R})+F^{\mu\nu}((f_{\sigma})_{R}P^{\sigma}%
\end{equation}
It then follows that the state $\omega$ cannot have sharp mass i.e. an
electrically charged particles is necessarily an infraparticle. Furthermore
the algebraic Lorentz symmetry is not unitarily implemented in a charged state
(spontaneous symmetry breaking) and even stronger: the momentum direction of a
asymptotically removed charged particle is a superselected quantity, since an
infraparticle is inexorably burdened by infinitely many infrared photons which
only can be pushed further into the infrared, but not eliminated.

That especially the last consequence appears strange to us is because our
intuition about charged states has been formed by charged particle in QM where
there is no superselection rule forbidding the coherent superposition of
electrons with different momenta. But as in real life, we cannot accept the
good parts of our most successful theory and reject structural consequences
which we do not like. Of course part of this difficulty lies in the
discrepancy between our structural knowledge and the present state of art
about perturbation theory. The following sections are dedicated to the problem
how to overcome these problems through a radical reformulation of gauge theory.

As mentioned on several occasions, interacting QFT is in many aspects much
more radical than QM. Using a picturesque metaphor one may say that it
realizes a benevolent form of Murphy's law: \textit{states which are not
interdicted by superselection rules to mutually couple, do inevitably couple}.
The culprit for this complication (or, depending one one's viewpoint, this
blessing) which contributes a fundamental aspect to QFT which is not shared by
QM, is the inexorable occurrence of the kind of vacuum polarization resulting
from the realization of the locality principle in the presence of an
interaction; the sharing of $\hbar$ does not bring them closer. So the
question arose whether the charged one-particle states can remain unaffected.
To exemplify what may happen, the idea of infraparticles was proposed in the
mathematically controllable context of soluble two-dimensional models for
which the anomalous power cut in the variable ($\kappa-m)$ of the
Kallen-Lehmann spectral function resembled the power behavior in the YFS work
for the soft photon inclusive cross section.

In the 70s there appeared the first $n^{th}$ order perturbative calculation
which addressed the infrared properties of charge-carrying fields; they could
be interpreted as confirming the momentum space infraparticle structure of
charged states \cite{Kibble}. But this did not close the issue since they had
two drawbacks; first they did not deal with gauge invariant charged fields and
second they had nothing to say about the possible weaker localization, which
in those 2-dimensional infraparticle models was the root of the infrared
problem. Some of the perturbative observation were subsequently derived in a
context which does not directly refer to perturbation theory. For example the
statement that there are no states with nontrivial charge nor states on which
the Maxwell equations hold on indefinite metric spaces (as the Gupta-Bleuler
or BRST settings) showed that the covariant perturbation theory involving
charged fields has very serious physical defects \cite{Fr-Str}\cite{Mo-Str}.
At best one can use the pointlike field formalism for the calculation of gauge
invariant vacuum expectation values and with the help of the Wightman
reconstruction theorem arrive at a new Hilbert physical Hilbert space and
physical (gauge-invariant) operators acting in it. Focussing attention on the
local observables, the problem of infraparticles and spontaneous breaking of
the Lorentz symmetry in charged states was taken up in \cite{F-Mo-Str} Further
insights came from comparing the quantum problem of localization of charges
with semiclassical arguments \cite{BDMRS}.

The most important structural (nonperturbative) enrichments, which came out of
the infrared problem and its roots in noncompact localization, are certainly
the aforementioned conclusions drawn from an appropriate formulation of the
quantum Gauss law in conjunction with a nontrivial charge which led to charged
states with an infinite extension. In the next section we will start to close
the large gap between structural insight and computational implementation.

\section{Localization peculiarities of zero mass Wigner representations}

In this section we will start to address some recent theoretical observations.

There is a very subtle aspect of modular localization which one encounters in
the second Wigner representation class of \textit{massless finite helicity
representations\footnote{There are 3 positive energy classes: massive,
massless with finite helicity and massless with infinite spin. The first ans
the third have the largest cardinality.}} (the photon-graviton class) which
recently attracted some attention \cite{MSY}. Whereas in the massive case all
spinorial fields $\Psi^{(A,\dot{B})}$ the relation of the physical spin $s$
with the two spinorial indices follows the naive angular momentum composition
rules \cite{Wein}%
\begin{align}
\left\vert A-\dot{B}\right\vert  &  \leq s\leq\left\vert A+\dot{B}\right\vert
,\text{ }m>0\label{line}\\
s  &  =\left\vert A-\dot{B}\right\vert ,~m=0 \label{line2}%
\end{align}
the "covariantization" of the zero mass finite spin representations leads to a
much stricter relation between the given physical helicity and the possible
dotted/undotted components of the covariant spinorial formalism for which the
Poincar\'{e} covariance can be extended to conformal invariance. For helicity
$s=1$ the best one can do is to work with covariant field strength $F_{\mu\nu
}$ which in the spinorial formalism correspond to wave functions $\Psi
^{(1,0)},$ $\Psi^{(0,1)};$ the desired vectorpotential representation
$\Psi^{(\frac{1}{2},\frac{1}{2})}$ corresponding to the classical
vectorpotential $A_{\mu}(x)$ simply does not occur in the Wigner
representation theory based the use of Hilbert space \footnote{The relation
between Wigner's one-particle representation theory and free fields will be
explained in more detail below. In fact there is a one-to-one correspondence
which permits to use the same letter for both (see below).}. Hilbert space
pisitivity is of no concern in classical field theory; A pointlike classical
field as $A_{\mu}(x)$ is a classical field as any other constraint classical
field i.e. as a member of an infinite dimensional subspace of a field space
subject to the classical BRST and/or Batalin-Vikovitski formalism.

Most particle theorists who access QFT by quantizing classical field
structures do not think much about why such a procedure leads inevitably to a
loss of Hilbert space (Gupta-Bleuler, BRST formalism) which is the basic
pillar of QT; they only take an indirect notice and often are not aware that
they have left the quantum physical terrain during the calculations to which
they must return at the end of the day in order to have physically
interpretable results. The reason is that quantum phenomena which have no
counterpart on the classical side are easily overlooked in a classical
parallelism as quantization; but they stick out very clearly in an intrinsic
approach to QFT as Wigner's representation theoretical setting.

At the root of this problem is a \textit{fundamental clash} between the
existence of pointlike covariant generating fields (or corresponding
generating wave functions in the Wigner representation space) and the
existence of spinorial fields which do not satisfy the restricted relation
(\ref{line2}). Let us for brevity introduce the following terminology: all
fields fulfilling this restricted relation will be called "field strength" and
the ones which cause that clash are "tensorpotentials". In the case of s=1 the
field strength $F_{\mu\nu}$ would be the field strength in Maxwell's theory
whereas the quantum theoretically incriminated $A_{\mu}(x)$ is the
vectorpotential; for s=2 the field strength is a 4-index tensor whose symmetry
properties are those of the Riemann tensor etc. The setting can be generalized
to half-integer s in which case s=3/2 (Rarita-Schwinger field) is the lowest
spin for which the clash of spinor potentials with the Hilbert space structure
occurs. It will be seen below that the tensor representation of the
representation type $D^{(\frac{s}{2},\frac{s}{2})}$ are the most interesting ones.

The resolution of the clash between the existence of pointlike potentials and
the Hilbert space of QT is very interesting. Instead of ceding on the side of
the Hilbert space, the clear message is to keep the latter but do not insist
in pointlike covariant localization but rather allow semiinfinite stringlike
generators\footnote{As a pointlike field is a generator (the
distribution-theoretical limit) of compact localized operator algebras, the
string-like generator serves the same purpose for noncompact regions in case
the pointlike generation is not possible-.}. In order to not be misunderstood,
all the $(m=0,s\geq1)$ Wigner representations possess pointlike generators but
their covariantization does not produce those covariant fields which one needs
for formulating renormalizable interactions.

As mentioned the standard path is to follow "quantum gauge theory" namely to
keep pointlike fields and the well-studied perturbative formalism at the prize
of an indefinite metric setting which usually comes with the introduction of
additional unphysical degrees of freedom (ghosts). Since in typical
perturbation calculations one does not use the Hilbert space norm for control
of convergence (as Schwartz inequality) there is no problem. At the end of the
calculation one has to reconvert the calculated correlation functions (e.g. by
cohomological arguments) into a Hilbert space setting; this is the step from
gauge variant to gauge invariant objects. To avoid any misunderstanding we are
not proposing a new string localized Hilbert space based approach because of
any suspicion that the gauge theoretical setting may be incorrect. Rather the
reason is that the latter is incomplete because the physically most important
charge carriers are not described within the standard gauge formalism; hence
looking for an alternative has not only a philosophical side but there are
also hard-core physical reasons. masks important physical properties for
example the noncompact localization of charges, not to mention even more
serious problems in case of Yang-Mills interactions (just those problems which
are connected to confinement and invisibility of states). All these nonlocal
(more precisely stringlike) effects must happen in the last step namely the
cohomological descent; this radical change from pointlike gauge variant to
stringlike gauge invariant overburdens the cohomological step apart from those
objects which remain pointlike which are the pointlike gauge invariants. The
most important physical objects as the charged fields remain outside the range
of this gauge formalism.

As will be seen keeping the Hilbert space and not imposing impossible
requirements on localization leads automatically to string-localized covariant
potentials $\Psi^{(A,\dot{B})}(x,e)$ which contain a spacelike string
direction and are localized on $x+\mathbb{R}_{+}e$ for all $(A,\dot{B})$ as in
(\ref{line}) which includes the covariant vectorpotential $A_{\mu}(x,e)$ for
s=1 and the tensorpotential $g_{\mu\nu}(x,e).$ The field strength for spin s
have scale dimension which s+1 (the ones with higher dimensions can be written
as derivatives) whereas the potentials have scale dimensions which fills the
space between 1and s+1, more precisely $1\leq dim\Psi^{(A,\dot{B})}(x,e)\leq
s.$ Of particular importance are the potentials $\Psi^{(\frac{s}{2},\frac
{s}{2})}(x,e)$ because their scale dimension is $\dim\Psi^{(\frac{s}{2}%
,\frac{s}{2})}=1$ independent of s; the previous fields $A_{\mu}(x,e)$ and
$g_{\mu\nu}(x,e)$ are examples. They lead to interactions which are
renormalizable in the sense of power counting. For pointlike fields
renormalizability in the sense of power counting is synonymous with
renormalizability whereas for string-localized fields this still must be established.

Since this clash between quantum theoretical positivity/unitarity and the
existence of pointlike generators with a prescribed covariant transformation
property is central to our new proposal, some more remarks are appropriate.
The problem of constructing covariant free fields from Wigner's unitary
representation theory of the Poincar\'{e} can be systematically solved in
terms of intertwiners $u(p,s)$ which are $\left(  2A+1\right)  \left(
2\dot{B}+1\right)  $ component functions on the mass shell (which then may be
rewritten into the tensor calculus). These intertwiners (between the unitary
and the covariant representation) and their adjoints can be systematically
computed, either by using group theoretic methods \cite{Wein} or "modular
localization" \cite{MSY}. For the massless finite helicity case with its
different "little group" most of the intertwiners do not exist (among them the
vectorpotential); only those for which the spinorial indices are related to
the physical spin in the more restrictive manner as in the second line
(\ref{line}) remain available. All the missing covariant realizations can
however be recovered if one allows the intertwiners to be dependent on a
string direction $u(p,e).$ This construction is most conveniently done in the
modular localization setting \cite{MSY}.

This construction has no counterpart in the classical Lagrangian setting and
hence cannot be formulated in the functional integral setting (perhaps one
reason why the string-localized potentials have only been noticed recently).
But fortunately perturbation is intrinsically defined in QFT and does not need
any crutches from the Lagrangian formalism; whether free fields are
Euler-Lagrange fields or not is of no relevance for the working of
perturbation theory \cite{Weinberg}. However the importance of causal locality
which is most visible in the Epstein-Glaser approach becomes even mor
important in passing from pointlike to stringlike fields.

The remaining question is then why does one need vectorpotentials if the field
strength wave functions generate already the whole Wigner space or, if in
interacting QED the quantum field strength together with the charged matter
fields generate the full Hilbert space and form an irreducible set of
operators in it ? The short answer is that one does not need it for the
description of QED but the string-localized potentials play a crucial role in
understanding the delocalization of the charges; without having the potentials
transfer their stringlike localization to the matter fields the nonlocality of
the latter remains a mystery. But even staying in the free Maxwell theory the
use of a pointlike vectorpotential together with Stokes law would lead to a
zero result for the Aharonov-Bohm effect whereas the string-localized
potentials leads to the correct effect.

Some of the answers can already be given in the free theory in terms of the
generalized Aharonov-Bohm effect (next section). The implementation of
renormalizable interactions adds additional reasons.

If one keeps the quantum (unitarity, Hilbert space, probabilities) aspects,
then the way out is to relax the pointlike localization which underlies the
Lagrangian quantization approach. Localization is important for the physical
interpretation and the derivation of scattering theory, but the localization
principle does not require to be realized in a pointlike generating manner,
i.e. whereas there is no lee-way on the side of the quantum theory
requirements, there is no principle in QFT which requires that a theory can be
point-like generated. The mathematical question about the tightest localized
generating wave function (technically: wave function-valued distribution)
which by smearing with test functions generate the full Wigner representation
space has been answered: all positive energy representations have semiinfinite
stringlike distribution-valued generators and the massive as well as the
massless finite helicity representations permit pointlike distribution-valued
generating wave functions (generalized field strengths).

Only the zero mass \textit{infinite} spin representation are intrinsically
string-localized in the sense that there is no "field strength" which
generates the same representation \cite{Jak}. In \cite{MSY} it was conjectured
that also the associated QFT has no pointlike generated subalgebras based on
the argument that their existence would lead to implausible consequences, but
the issue is not completely settled on the level of mathematical
physics\footnote{In particular since there has been a recent claim (without
proof) to the contrary by Ch. K\"{o}hler, Institut fuer Theoretische Physik,
University Goettingen, work in progress.}. there is also the aforementioned
curious difference between the two pointlike generated representation, which
at closer inspection reveals a subtle distinction in localization aspects. In
the massive case the little group is compact, whereas in the massless case it
is the noncompact Euclidean stability group $E(2)$ of a lightlike direction.
In the finite helicity case this representation is finite dimensional, hence
necessarily a unfaithful (degenerate) representation. Only in Wigner's
infinite spin family this noncompact stability group ("little group") is
faithfully represented.

All three families of representations massive, massless finite helicity and
massless infinite spin are positive energy representations and there is a
structural theorem \cite{BGL}, stating that all unitary positive energy
representations of the Poincar\'{e} group (irreducible or not) can be
generated by semiinfinite string-localized fields. But only in case of the
infinite spin family (infinite dimensional representation of the stability
group) this is the best possible localization. For the other two families the
best (sharpest localized) generators are pointlike fields (operator valued
distributions in the associated free field theory) which makes them accessible
via the classical-quantum parallelism known as quantization.

The above observation about the existence of gaps in the spinorial covariance
spectrum (\ref{line}) means that even though both families of spinorial
representations are finite dimensional, the noncompactness of the little group
still makes its presence felt by not allowing most of the spinorial generators
which occur in the massive case. Using a terminology which generalizes the
(m=0, s=1) case of (noninteracting) electromagnetism, we may talk about
pointlike "field strengths" and their associated string-like \ "potentials ",
which taken together \textit{reconstitute the full spinorial spectrum} in the
first line of (\ref{line}). So from now on "field strengths" denote the
covariant pointlike objects the second line of (\ref{line}), whereas
"potential" is the generic terminology for the string-localized remainder
which recover the full spinorial spectrum of the first line.

The main idea is of course that, although string localized vectorpotentials do
not fit into the standard formalism, it is better to face the problem on its
physical side; always with the increased awareness that localization is the
dominant principle, and in order to uphold it rather change the formalism than
to compromise on physical principles. In order to avoid being misunderstood,
we are not criticizing the gauge theory formalism in its efficiency to deal
with local observables and we even have some sympathy for the temporary
trespassing of the most cherished principles of QT by ghosts in the name of
computational efficiency. Without the contributions of Veltman and t'Hooft,
Faddeev and Popov and the BRST formalism of Stora et al., a consistent
extension of the renormalization setting of QED to the standard model would
not have been possible. But meanwhile almost 40 years have past, and although
most people agree that the theory is nowhere near its closure, nothing of
conceptual significance has happened. It is natural in such a situation to
search for unexplored corners of QFT and the issue of localization, which is
central to the present work, is certainly a rather dark corner even in QED.
The attempt to complete a theory may still turn into a unexpected radical change.

This paper is a plea to follow the localization principles and develop a new
string-localization compatible formalism. We will present some of the first
steps in this direction. It is worthwhile to mention already here, that the
origin of the string-localized electric charge-carrying fields in QED,
including their infrared aspects, is the result of the interaction of the
matter current with the stringlike vectorpotentials. Whereas the influence of
the stringlike localization on their own physical properties in
selfinteracting Yang-Mills models is hard to foresee without understanding
their infrared aspects, it is clear that the localization of the linearly
related field strength in abelian gauge theories remain pointlike and that
although the physical content of QED can be described in terms of physical
string-localized matter fields $\psi(x,e)$ and pointlike $F_{\mu\nu}(x),$ the
stringlike vectorpotentials remain indispensable in the formulation of the
interaction and the perturbative calculations.

Since such perturbative calculations are fairly involved and the cohomological
descent from gauge-variant to gauge-invariant correlation is only simple for
correlations involving $F_{\mu\nu}(x)$ but not $\psi(x,e)$, it is not without
interest that the quantum version of the Aharonov-Bohm like (next
section)\footnote{There is a curious analogy between the abandonement of the
ether, which was mainly important for the post Maxwell-Lorentz development of
particle physics, and the string-localization of potentials which only has
severe consequences for the charged sector in the interacting theory.} permits
to see the important role of the de-localized vectorpotential.

It is an ineradicable prejudice to believe that perturbation theory has to be
set up in terms of quantized Lagrangians and functional integrals. The
conceptually most pleasing perturbative approach consists in coupling
covariant free fields (obtained as above by covariantizing the 3 classes of
positive energy Wigner representations) together in form of invariant
polynomial interactions\footnote{One may call it the interaction Lagrangian
but the free fields/potentials used may and generally will not have a
Euler-Lagrange structure. That perturbation theory does not require the
existence of a Lagrangian for a free field was already known to Weinberg
\cite{Weinberg}.} ("causal perturbation theory") which only contain pointlike
fields and stringlike potentials of short distance scaling dimension
$d_{sca}=1,$ since any higher dimensional fields/potentials would violate the
power-counting requirement. Whether the coupled free fields obey an
Euler-Lagrange equation is irrelevant for this perturbation theory, a fact
which was already known to Weinberg \cite{Weinberg}.

It is another equally ineradicable misconception (usually related to the
previous one), that the perturbative treatment of QFT contains intrinsic
infinities which are "renormalized away". Fact is that certain implementations
of perturbation theory, especially those which treat the problem in the
quantum mechanical spirit of bringing certain operator functions of free field
under control without paying much attention to the fact that fields, even in
the absence of interactions, are rather singular objects which want their
mathematical role as operator-valued distributions to be taken serious. If one
does not do this, one still has a chance, but there is a prize to be paid: the
removal of cutoff infinities (or ad hoc regularization parameters). But there
is always a finite way to arrive at those results, and if this would not be
so, the whole renormalization approach would not have credibility.

For $s\geq1$ such potentials with $d_{sca}=1$ exist for all helicities. The
property which is crucial in this approach is the localization structure, in
fact the intertwiner formalism which leads to the spinorial fields
(\ref{line}) can be solely based on modular localization instead of group
theory \cite{MSY}. It is therefore not surprising that also the
renormalization procedure can be built on the iterative fulfillment of the
localization principle combined with a requirement of keeping the scaling
degree of the counter-terms at its minimal possible value: the string-extended
Epstein-Glaser approach.

For completeness and also for making some subsequent speculative remarks more
comprehensible, it is important to say something about the
large\footnote{Although it is a zero mass representation, the faithfulness of
the representation of the little group brings a continuous parameter into the
game which leads to a higher cardinality of representations than in the finite
spin case.} third family of \textit{infinite spin representations}. These are
irreducible massless representations in which the euclidean E(2) stability
subgroup is faithfully represented. In this case the representation theory of
the little group leads to an infinite dimensional Hilbert space \cite{MSY}.
The Casimir invariant of the little group (i.e. the E(2) analog of the mass
operator) takes on continuous values. In order to avoid the somewhat
misleading terminology "continuous spin" in the older literature associated
with a continuous representation of the "little" Hilbert space, it may be more
appropriate to follow the recent terminology and refer to the "infinite spin representations".

Before we return to the discussion of consequences of stringlike localization,
it is helpful to formalize the covariant fields for all three families. Some
of these formulas can be found in the first volume of Weinberg's book
\cite{Wein} e.g. the following formula for massive free fields
\begin{equation}
\Psi^{(A,\dot{B})}(x)=\frac{1}{\left(  2\pi\right)  ^{\frac{3}{2}}}%
\int(e^{-ipx}u^{(A,\dot{B})}(p)\cdot a(p)+e^{ipx}v^{(A,\dot{B})}(p)\cdot
b^{\ast}(p))\frac{d^{3}p}{2\omega}%
\end{equation}
where the dot stands for the sum over 2s+1 spin component values. The
operators $a^{\#}$ and $b^{\#}$ are the momentum space annihilation/creation
operators which transform according to the respective Wigner representation.
The intertwiner $u^{(A,\dot{B})}(p,s_{3})$ and their charge conjugate
counterpart $v$ convert the Wigner representation into the covariant
representation. They are rectangular matrices transforming a $2s+1$ component
Wigner spin into a $(2A+1)(2\dot{B}+1)$ component covariant space. For a given
physical spin s there is an infinity of possibilities of which one only uses
the ones with low $A,\dot{B}$ values which happen to have the lowest scaling
degrees. The generating wave functions of the beginning of this section are
obtained by replacing the Wigner creation/annihilation operators by the
function $f(p)=1.$

As mentioned, Weinberg's method to compute these intertwiners was group
theoretical, but one can also base the computation on modular localization
\cite{MSY}; this is not surprising since covariance and locality of states are
closely linked.

Practically the same formula, with the only change in the range of $s_{3}=\pm
s$ and different expression for the intertwiners, holds for the massless case.
However there is an important caveat, the formula exists only for the
restricted $A,\dot{B}$ values in the second line of (\ref{line}) i.e. only for
field strengths.

If one allows string-localization one can recover all the lost spinorial
representations. These "potentials" are (by definition) all string-localized
and obey for $\left\vert A+\dot{B}\right\vert \geq s\geq\left\vert A-\dot
{B}\right\vert ,$ (field strengths excluded i.e. $s\neq\left\vert A-\dot
{B}\right\vert )$ the following formula%

\begin{align}
&  \Psi^{(A,\dot{B})}(x;e)=\frac{1}{\left(  2\pi\right)  ^{\frac{3}{2}}}%
\int(e^{-ipx}u^{(A,\dot{B})}(p,e)\cdot a(p)+e^{ipx}v^{(A,\dot{B})}%
(p,e)b^{\ast}(p))\frac{d^{3}p}{2\omega}\\
&  A^{\mu}(x,e)=\frac{1}{\left(  2\pi\right)  ^{\frac{3}{2}}}\int
(e^{-ipx}u^{\mu}(p,e)\cdot a(p)+e^{ipx}\overline{u^{\mu}(p,s_{3};e)}\cdot
a^{\ast}(p))\frac{d^{3}p}{2\omega}\ \ \ \ \ \ \ \ \ \ \ \ \ \ \ \ \ \ ~~
\label{vector}%
\end{align}
where the dots stand for the sum over the two helicities $s_{3}=\pm s~$and the
vectorpotential intertwiner, which has been written down separately, has the
form%
\begin{equation}
u^{\mu}(p,s_{3};e)_{\pm}=\frac{i}{pe+i\varepsilon}\{(\hat{e}_{\mp}(p)e)p^{\mu
}-(pe)\hat{e}_{\mp}^{\mu}(p)\}
\end{equation}
and the $\hat{e}_{\pm}$ denotes the two photon polarization vectors to be
distinguished from the string direction. These operators transform covariantly
and have stringlike commutation relations%

\begin{align}
&  U(\Lambda)\Psi^{(A,\dot{B})}(x,e)U^{\ast}(\Lambda)=D^{(A,\dot{B})}%
(\Lambda^{-1})\Psi^{(A,\dot{B})}(\Lambda x,\Lambda e)\\
&  \left[  \Psi^{(A,\dot{B})}(x,e),\Psi^{(A^{\prime},\dot{B}^{\prime}%
)}(x^{\prime},e^{\prime}\right]  _{\pm}=0,~x+\mathbb{R}_{+}e><x^{\prime
}+\mathbb{R}_{+}e^{\prime}\nonumber
\end{align}

As expected, the scaling degree of the potential is $d_{sca}(A^{\mu}(x,e))=1$
i.e. better than that of the field strength. The resulting two-point function
is of the form \cite{JRio}%

\begin{align}
&  \left\langle A_{\mu}(x;e)A_{v}(x^{\prime};e^{\prime})\right\rangle =\int
e^{-ip(x-x^{\prime})}W_{\mu\nu}(p;e,e^{\prime})\frac{d^{3}p}{2p_{0}}%
\label{tp}\\
&  W_{\mu\nu}(p;e,e^{\prime})=-g_{\mu\nu}-\frac{p_{\mu}p_{\nu}}{(p\cdot
e-i\varepsilon)(p\cdot e^{\prime}+i\varepsilon)}(e\cdot e^{\prime
})+\nonumber\\
&  +\frac{p_{\mu}e_{\nu}}{(e\cdot p-i\varepsilon)}+\frac{p_{\nu}e_{\mu
}^{\prime}}{(e^{\prime}\cdot p+i\varepsilon)}\nonumber
\end{align}
The presence of the last 3 terms is crucial for the Hilbert space structure;
without them one would fall back to the indefinite metric and negative probabilities.

Either from the two-point function or more directly from the form of the
intertwiners one reads off the following two relations:%

\begin{equation}
\partial_{\mu}A^{\mu}(x,e)=0=~e_{\mu}A^{\mu}(x,e) \label{Lor}%
\end{equation}
These formulas are not imposed but are consequences of the requirement of
having a covariant vectorpotential in the Wigner Hilbert space which, since it
cannot be point-local, brings in an additional geometric parameter $e.$

Note that the string-localized potential looks like the axial gauge potential
in the gauge theoretical setting, where $e$ is a gauge parameter which,
different from the above definition, is inert against Lorentz transformations.

The difference between the gauge interpretation and string-localization is
more conceptual than formal. The fluctuation in both parameters $x,e$ ($e$~a
point in 3-dim. de Sitter space) is very important for the lowering of the
$x$-dimension, namely $d_{x}=1,$ at the cost of fluctuation in $e$ which
manifest themselves as string-caused infrared divergences; the latter explain
why the axial gauge, which overlooks the fluctuations in $e$ is intractable
(it was never of any computational use); despite of its welcome Hilbert space
structure, the fluctuations in $e$ prevented to control these divergencies, in
fact it was not even possible to understand their physical origin. The
localization based approach on the other hand treats the string potentials as
distributions in $x$ \textit{and} $e$ and addresses the question if and how
the different $e_{i}$ have to coalesce at the end. The axial gauge is besides
the Coulomb gauge the only one which can be accommodated in a physical Hilbert space.

The interpretation of the $e$ as a fluctuating covariant string direction
rather than as fixed gauge parameter in the interpretation as a
string-localized potential is the only meaningful interpretation. Hence the
standard argument in favor of gauge invariance based on returning to a
physical space from the unphysical indefinite metric formulation has lost its
conceptual basis. But a new problem has emerged, namely how to treat
fluctuating string directions. If one has grown up with a "gauge principle",
this may seem surprising, but the surprise should not be new, one could have
asked the crucial question "does the axial gauge with its infrared
divergencies (even for off-shell expectations of matter fields) fit into the
standard gauge-ideology ?" already a long time ago. The answer to this
question is that it does not, it is really a Hilbert space description in
which the pointlike was changed with semiinfinite stringlike localization; but
it required the use of the modular localization concept \cite{BGL}\cite{MSY}
to raise the awareness about this issue.

In order to obtain a theory in which the interaction between the
vectorpotentials and matter leads to a subalgebra of local observables, one
needs a relation which connects the potential for two different directions
\begin{align}
&  A^{\mu}(x,e)\rightarrow A^{\mu}(x,e^{\prime})+\partial^{\mu}\Phi
(x;e,e^{\prime})\label{change}\\
&  \Phi(e,e^{\prime};x)=\int e_{\mu}A^{\mu}(x+te^{\prime},e)dt\nonumber
\end{align}
The proof of pointlike locality of certain fields amounts to the e-indepence;
this is not different from the proof of independence of gauge parameters in
the standard gauge theoretical setting. But the main purpose of this formalism
is not the identification of local observables and the calculation of their
correlation functions but rather to incorporate the string-localized charged
fields and their infraparticles into the perturbative formalism.

In fact the string localization suggests to view the change of $e$ as the
result of two subsequent Poincar\'{e} transformations a Lorentz rotation
$\Lambda(e,e^{\prime})$ around the origin which transforms $e$ into
$e^{\prime}$ and a conjugation by a translation, hence together%
\begin{align}
&  A(x,e^{\prime})=V(e,e^{\prime};x)A(x,e)V(e,e^{\prime};x)^{\ast
},~V(e,e^{\prime};x)\equiv U(x)U(\Lambda(e,e^{\prime}))U(-x)\\
&  V(e,e^{\prime};x)\varphi^{\ast}(x)\partial_{\mu}\varphi(x)V(e,e^{\prime
};x)^{\ast}-\varphi^{\ast}(x)\partial_{\mu}\varphi(x)=\partial_{\mu}%
\Phi(e,e^{\prime};x)\nonumber
\end{align}
which leads to an interpretation of $\Phi$ in terms of spacetime operations.
This in turn suggests that in a theory in which the stringlike vectorpotential
interacts with matter fields as in QED, \textit{composite charge neutral
operators} which contain derivatives acting between charge-carrying operators
as $\varphi^{\ast}\partial_{\mu}\varphi$ should "feel" the $e\rightarrow
e^{\prime}$ change by producing additive $\partial_{\mu}\Phi$ terms as in the
second line above, so that gauge invariance formally corresponds to string
independence in the new setting. In view of the fact electrically charged
fields cannot be better than string-localized and that the Lorentz invariance
is spontaneously broken in charged sectors one expects that an operational
description of the $e$ change may connect these two facts within a
perturbative setting. In that case one would have three kinds of behavior
under change of $e$: $e$-independent i.e. pointlike localized operators
corresponding to gauge invariant observables, operators which change
additively under changes of $e$ similar to a change under a symmetry
transformation and operators on which these symmetry-like transformations are
spontaneously broken. This description, if possible, would strengthen the old
observations of an interrelation of gauge and spacetime properties
(Lorentz-covariance and localization) and could reconcile rigorous structural
observations as the spontaneous breaking of the Lorentz symmetry in
electrically charged sector of the theory with the formalism of perturbation
theory for string-localized objects. In case of selfinteracting vector- or
tensor-potentials, as in Yang Mills and Einstein-Hilbert interactions, the
changes under $e$ in the interacting theory are not the same as for free
fields. All these problems, which are connected with string-localization will
be left to future work.

A closely related question is whether Hertz-potentials, which were intoduced
by Hertz into Maxwell's theory as useful computational tools, have also a
beneficial role to play in QED. Formally they would be described by
string-localized antisymmetric tensor fields with zero scale dimension so that
derivatives and exponentials have positive dimensions. They appear in the work
of Penrose in connection with asymptotic behavior of zero mass higher spin
fields \cite{Pen}.

Another special case of significant interest is the case of s=2 whose field
strength with the lowest scale dimension is an object $R_{\mu\nu\kappa\lambda
}$ with the linear properties of the Riemann tensor and an $d_{sca}=3$ and a
string-like potential $g_{\mu\nu}(x,e)$%
\begin{align}
\left\langle g_{\mu\nu}(x,e)g_{\kappa\lambda}(x^{\prime},e^{\prime
})\right\rangle  &  =\int\frac{d^{3}p}{2p_{0}}e^{-ip(x-x^{\prime})}W_{\mu
\nu\kappa\lambda}(p;e,e^{\prime})\\
W_{\mu\nu\kappa\lambda}(p;e,e^{\prime})  &  =W_{\mu\alpha\nu\beta\kappa
\rho\lambda\sigma}^{R}\frac{e^{\alpha}e^{\beta}e^{\prime\rho}e^{\prime\sigma}%
}{(e\cdot p-i\varepsilon)^{2}(e^{\prime}\cdot p+i\varepsilon)^{2}}\nonumber
\end{align}
where the superscript $R$ refers to the field strength 2-point function whose
8 tensor indices correspond to the 4 tensor indices of the independent field
strength and reflect the fact that the relation between the potential and the
field strength is the linearized version of that between the metric tensor and
the Riemann tensor \cite{JRio}. The $e$-dependent factor obviously improves
the short distance properties. As expected the $g_{\mu\nu}$-potential has
$d_{sca}=1$.

String-localized potentials with $d_{sca}=1$ can also be constructed for
massive theories, even though there is no compelling reason from the viewpoint
of the Wigner representation theory for doing this since pointlike fields with
a fixed physical spin exist for all spinorial pairs fulfilling (\ref{line}).

In this case the \textit{only reason would be the power counting requirement}.
Since the increase of the short distance dimension with spin happens
independent of the presence of a mass, there would be no renormalizable
interaction of a spin one massive $A_{\mu}(x)$ field with other s=0 or
s=$\frac{1}{2}$ matter fields, the only way out is to take a string-localized
massive $A_{\mu}(x,e)$ with $d_{sca}=1$ instead of $d_{sca}=2$ for $A_{\mu
}(x).$ This enlarges the number of candidates for renormalizable interactions
from a finite number to infinitely many. But even if some interactions which
are power counting renormalizable turn out to lead to mathematical consistent
theories, unless they have local observables in the form of pointlike
generated subtheories, they are physically unattractive.

For infinite spin one finds \cite{MSY}\cite{JRio}%
\begin{align}
&  \Psi(x;e)=\frac{1}{\left(  2\pi\right)  ^{\frac{3}{2}}}\int(e^{-ipx}%
u(p;e)\circ a(p)+e^{ipx}\sum_{s_{3}=\pm s}v(p;e)\circ b^{\ast}(p;e))\frac
{d^{3}p}{2\omega}\nonumber\\
&  u(p;e)(\kappa)=\int_{\mathbb{R}^{2}}d^{2}ze^{ikz}(\xi(z)B_{p}%
^{-1}e)^{-1+i\alpha} \label{inf}%
\end{align}
where $B_{p}$ is the $p$-dependent family of Lorentz-transformation, selected
in such a way that they transform the reference $\bar{p}$ on the irreducible
orbit into the generic $p,$ and the circle product stands for the inner
product in the "little" Hilbert space which consists of square integrable
functions of a two-dimensional Euclidean space $\bar{f}\circ g=\int\bar
{f}(\kappa)g(\kappa)d^{2}\kappa.$ So the Wigner $a(p)^{\#}$ operators and the
intertwiners depend on the euclidean $\kappa$ variables make the dependence on
e much more involved than in the finite helicity case. Nevertheless they are
string-localized for all values of $\alpha.$ The $\kappa$ dependence of the
intertwiners results in a stronger form of string localization than for zero
mass potentials.

This stronger form of delocalization shows up in the fact that there are no
pointlike field strengths \cite{Jak}. In fact the operator algebra generated
by these fields apparently has no compactly generated observable (pointlike
generated) subalgebras\footnote{The pedestrian argument for bilinear operators
in \cite{MSY} can be generalized to monomials of arbitrary orders, but an
elegant proof based on modular methods is still missing.}. In general
stringlike generators of algebras which cannot be assembled as a union from
compact parts (as our favorite example of operators carrying a nontrivial
electric charge) do generate reducible states under the action of the
Poincar\'{e} group if applied to the vacuum which are pointlike generated; the
localization of operator algebras and the localization of states are two
different pairs of shoes. But representations of the Poincar\'{e} group which
have infinite spin components in their reduction can only appear in operator
algebraic structures which were string-localized.

The only states which are intrinsically string-localized are those associated
to the infinite spin representation. Theories in which they occur would have
serious problems with being accessible to observations\footnote{This may have
been the reason why Weinberg \cite{Wein} dismissed them as unphysical, despite
their fulfillment of the positive energy requirement.}. This is because
"counters" in QFT are compactly localizable or, in order to avoid vacuum
polarization problems \cite{Haag}, they should be at least localizable in the
sense of quasilocality. Such a counter cannot register an intrinsically
string-localized state, so that quantum matter related to the third Wigner
class remains "invisible" despite the fact that it carries nonvanishing
energy-momentum (and hence susceptible to gravity). Theories containing such
representations are candidates for "invisible" quantum matter. So maybe
Weinberg's "no" at the time of writing his book should be weakened to "not yet".

\section{Aharonov-Bohm effect for and violation of Haag duality}

Suppose one generates the ($m=0,s\geqslant1$) Wigner representation space (or
the associated net of local algebras in the Wigner-Fock space) with pointlike
field strength wave functions. Does the theory let us know that we forgot that
there are string-localized potential which want to play an important physical
role? In this case the Wigner representation should signal by its localization
properties that there is a difference in localization between the massive to
the massless representation, but does it really do this? Is there an intrinsic
difference within the Wigner representation with respect to the modular
localization structure which goes beyond the covariantization formula
(\ref{line})?

There is indeed a subtle representation theoretical distinction which is
connected with \textit{Haag duality}. Whereas for simply connected convex
double cone regions the localization spaces (real subspaces of the Wigner
representation space, which are defined in terms of modular
localization\footnote{The K-spaces are real subspaces of the complex Wigner
space which are defined as eigenspaces of the involutive Tomita S-operator.
For a presentation of these spaces which is close to the spirit of the present
paper we refer to \cite{interface1}.} (\ref{S})) one finds Haag duality for
both the massive as well as the massless finite spin case%
\begin{align}
K(\mathcal{O}^{\prime})  &  =K(\mathcal{O})^{\prime}\text{ }or~K(\mathcal{O}%
)=K(\mathcal{O}^{\prime})^{\prime}\label{1}\\
\mathcal{A(O})  &  =\mathcal{A(O}^{\prime})^{\prime},~~\mathcal{O}%
~double\text{ }cone \label{2}%
\end{align}
but there are interesting differences for \textit{non simply connected
regions} as relative causal complements of a smaller spacetime double cone
within a larger one (the causal completion of a torus) where (\ref{1}) refers
to the Wigner representation theory and (\ref{2}) is the second quantized
algebraic version. The interpretation of Haag duality in the standard (von
Neumann) quantum theoretical setting of measurements is that any measurement
which is compatible with all measurement inside a causally complete spacetime
region, must be associated with the causally disjoint observables (\ref{2});
this rule can only be broken for very special and important reasons.

One knows from some old (but unfortunately unpublished) work \cite{LRT} that
for free QED, i.e. for the (m=0,s=1) Wigner representation, the Haag duality
breaks down for a spacetime region $\mathcal{T}$ ~which results from the
causal complement of a double cone inside a larger double cone or by sweeping
an $x-t$ two-dimensional double cone subtended from a spatial interval
$x\in\lbrack a,b],$ $0<a<b,$ by higher dimensional rotation around the origin.
In d=1+2 the result would be topologically equivalent to the inside of a
torus, whereas in d=1+3 it is the doubly connected 4-dimensional analog namely
the causal completion of a 3-dimensional torus region at a fixed time. One
then finds \cite{LRT}
\begin{align}
K(\mathcal{T}^{\prime})  &  \varsubsetneq K(\mathcal{T})^{\prime}\label{K}\\
or~K(\mathcal{T})  &  \varsubsetneq K(\mathcal{T}^{\prime})^{\prime
},~~\mathcal{A(T})\varsubsetneq\mathcal{A(T}^{\prime})^{\prime}%
\end{align}
where in the second line we also wrote the violation of Haag duality relation
in the interaction-free algebraic setting which are functorially related to
the one formulated in terms of modular localized real particle subspaces $K$
of the complex Wigner representation spaces.

These proper containment relations in the special case of the free
electromagnetic field are quantum field theoretic analogs of the
quasiclassical Aharonov-Bohm (also Ehrenberg-Siday) effect. This is an effect
of a localized classical magnetic flux in an infinitely long solenoid exerts
on a quantum mechanical electron which scatters on the solenoid even though it
stays outside its thin magnetic flux tube. Its quantum field theoretic
counterpart (A-B effect in QFT\footnote{Usually the A-B effect refers to the
semiclassical scattering \ of charged pstzivles on a thin solonoid whose
magnectic field remains inside.}) is more stringent; it states that despite
the continued validity of Stokes theorem for the quantum magnetic field, the
quantum magnetic potential has modular localization properties which differ
from the classical intuition since classically it should be localized on (or
at least near) the Stokes boundary circumference. In other words the effect
would disappear if the magnetic potential would be a pointlike field as in
gauge theory.

This does of course not mean that gauge theory is misleading, it is only a
warning against its unconstrained use. Matters of localization should only be
discussed in a Hilbert space i.e. \textit{after }having implemented the
invariance under BRST transformations. But since the BRST "symmetry" acts in a
nonlinear way, the construction of BRST invariant local correlations is very
difficult and for this reason rarely done; it is especially the case if
nontrivial Maxwell or Yang Mills charges lead to stringlike localizations
after having done the cohomological BRST descent.

There are two remedies for the A-B effect at hand, either one introduces the
notion of quantum cohomology of field strength \cite{LRT}, or one uses the
\textit{string-localized quantum vectorpotentials}. For free electromagnetic
field the first choice is completely adequate, but in the presence of
interaction only the formulation based on string-localized potentials has a
good chance to permit the perturbative construction of physical charged fields
and to really explain the origin of their de-localized nature.

In the cited work \cite{LRT} the connection with the A-B effect was not
mentioned and the representation theoretical basis of modular localization via
intersection of wedges was not yet available. The calculation was done in the
covariant field strength formalism of $\vec{E},\vec{H}$. The main purpose was
to formulate a warning against the use of pointlike vectorpotentials in QFT
which leads to a contradiction with the A-B effect and its extension i.e. the
violation of Haag duality (\ref{K}). This, and the remark that
string-localized magnetic potentials avoid this contradiction, is the precise
reason why we revisited this age old problem. Since in the massive case the
equality sign in (\ref{K}) continues to hold, the toroidal Haag duality
violation is the looked-for \textit{intrinsic representation theoretic
distinction} between massive and finite helicity massless representations.

Perhaps the best way to present this result is to say that the pointlike
localization for massless vectorpotentials clashes with the Hilbert space
positivity and since the latter is the essence of quantum theory, it is the
pointlike localization which has to cede. The only generalization of pointlike
localization turns out to be semiinfinite stringlike; the positive energy
condition of unitary representations of the Poincare group does not require to
introduce \textit{generating} wave functions (or free fields) which are weaker
localized than a semiinfinite string.

Following LTR one looks at a situation of two spatially separated, but
interlocking regions $\mathcal{T}_{1}$ and $\mathcal{T}_{2}$ in which one
represents as the smoothened boundary of two orthogonal unit discs $D_{1}$ and
$D_{2}~$which intersect in such a way that the boundary of one passes through
the center of the other. The delta function fluxes through the $D_{i}$ are
smoothened by convoluting $\star$ with a smooth function $\rho_{i}%
(\mathbf{x})$ supported in an $\varepsilon$-ball $B_{\varepsilon}$; the
interlocking $\mathcal{T}_{i}$ are then simply obtained as $\mathcal{T}%
_{i}=\partial D_{i}+B_{\varepsilon}~i=1,2.$ One computes the following
objects
\begin{align}
&  Im(e(\vec{g}_{1}),h(\vec{g}_{2}))\simeq\lbrack\vec{E}(\vec{g}_{1})\vec
{H}(\vec{g}_{2})]=\int\vec{g}_{1}(x)rot\vec{g}_{2}(x)d^{3}x=\label{alg}\\
&  =\int\rho_{1}(x)d^{3}x\int\rho_{2}(y)d^{3}y,~~\ \vec{g}_{i}=\vec{\Phi}%
_{i}\star\rho,~\vec{\Phi}_{i}(\vec{f})=\int_{D_{i}}\vec{f}d\vec{D}%
_{i}\nonumber
\end{align}
where we have written the result in two different ways, on the right hand side
the algebraic (commutator) expression and on the left hand side in terms of
the associated Wigner wavefunction. The $\Phi_{i}$ is the functional which
describe the flux through $D_{i}$, a kind of surface delta function.

The calculation of wavefunction inner product and its associated symplectic
form defined by its Imaginary part is more lengthy but straightforward. It is
needed because it contains the information of the modular localization. We
will omit it here and simply state the result. It confirms relation (\ref{K})
since $e(\vec{g}_{1})\in K(\mathcal{T}_{1}^{\prime})^{\prime},~h(\vec{g}%
_{2})\in K(\mathcal{T}_{2}^{\prime})^{\prime}$ but none of these two wave
functions are in the smaller spaces $K(\mathcal{T}_{i})$ since the algebraic
right hand side (\ref{alg}) is definitely nonvanishing. The QFT A-B effect is
the only known violation of Haag duality for which the duality violating
operators cannot be used for a "Haag dualization" i.e. an extension process by
which Haag duality can be recovered for the extended not of local algebras.

The calculation can be done entirely in terms of field strengths, there is no
need to use potentials and their two point functions (\ref{tp}). The first
term in (\ref{tp}) which contains only $g_{\mu\nu}$ and no string dependence
$e$ would lead to an indefinite inner product if taken for the two-point
function of vector potentials; in fact this would describe the indefinite
two-point function of pointlike vector potentials in the covariant Feynman
gauge; but restricted to the field strength it is perfectly positive. On the
other hand the full two-point function (\ref{tp}) is positive, this was the
main achievement of reconciling modular localization and positivity via
string-localization. For the cohomological argument supporting the QFT A-B
effect or breakdown of Haag duality, one does not need the potentials. It is
only if one wants to have a more operational argument for the discrepancy
between Stokes theorem and modular localization than that based on cohomology
that one needs the free stringlike potentials.

However the operational formulation in terms of \textit{string-localized
potential become absolutely crucial in the presence of interactions} for the
understanding of the properties of physical charges. I know of no
cohomological argument in terms of which one can understand the localization
properties of interacting Maxwell charges. The fact that the A-B effect
disappears if one uses the pointlike (and hence non Hilbert space) magnetic
potential $\vec{A}(x)$ in Stokes theorem shows that one has to be very careful
in drawing physical conclusions from the standard gauge formalism. Only after
the imposition of BRST invariance and the cohomological descent (too difficult
for correlation of charged field, only stated but never performed) has one
left the slippery ground.

There is no all-clear in the context of interacting gauge theories. It is
well-kown that there is no electric charge in a formulation QED in which a
pointlike vectorpotential is still present \cite{Fr-Str}\cite{F-Mo-Str}. In
such a case there exists only the nontrivial kinematical Dirac charge, but the
Maxwell charge vanishes. As in the case of the above Aharonov-Bohm effect,
this unphysical aspect disappears upon using the string-localized vector
potentials in the Hilbert space formulation. Although we will not go into
higher spin problems it may be interesting to remark on the side that the
string-localization in Hilbert space has extensions to higher spin potentials
whose scale dimension does not increase with s (with renormalizable
interactions in the sense of power counting) whereas an extension of the gauge
formulation beyond s=1 is not known.

The violation of Haag duality for conformal QFT on multifold connected
spacetime regions is part of modular theory and this raises the question
whether one can compute the modular group. The answer is positive and quite
interesting; it will be deferred to an appendix.

Some more remarks about (the algebraic) Haag duality and its breaking are in
order. Its validity for simply connected regions $\mathcal{O}$ (\ref{2})
defines a "perfect" world in which the quantum counterpart of the classical
Cauchy propagation holds i.e. a local algebra is equal to that of its causal
completion and the commutant of the algebra localized in the causal disjoint
$\mathcal{O}^{\prime}$ is equal to the original algebra.

As mentioned, interesting situations arise when the world of local quantum
physics is not perfect and the Haag duality is violated i.e. the right hand
algebra is genuinely bigger than the left hand side. The most common violation
results from an observable algebra which is localized in several disconnected
spacelike separated double cones (separated intervals in the chiral conformal
case \cite{Mue}). In case the observable algebra possess localizable
superselected charges, the right hand side for such a multi-disconnected
region $\mathcal{A}(\mathcal{O}^{\prime})^{\prime}$ is genuinely bigger
because the \textit{charge transporters} which carry the charge from one to
the other region are in $\mathcal{A}(\mathcal{O}^{\prime})^{\prime}$ but not
in $\mathcal{A(O});$ the charge transporters are globally neutral, but they
change the localization of charges between the localization regions.

Such models fall into the range of the DHR superselection theory \cite{Haag}.
The final result of this theory is the (unique) existence if a "field
algebra", which contains all superselected charges and a compact symmetry
group\footnote{The appearance of compact group theory via the localization
properties of observable algebras is perhaps the most surprising aspect of the
power of quantum localization \cite{DR}.}, which acts on the field algebra in
such a way that the observable algebra re-emerges as the fixed point
subalgebra. In the chiral case one can even compute geometric modular groups
for such situations. They are associated with higher diffeomorphism groups
beyond the Moebius group \cite{LMR} and they require to trade the standard
vacuum with the so-called split vacuum. In all those cases the violation of
Haag-duality is an indicator of the presence of charge superselection sectors
and a global symmetry. In an appendix the reader finds an interesting
illustration of such a situation.

The idea underlying the relation between the charge neutral observables and
the charge carrying fields can be best by borrowing a famous phrase from Marc
Kac in conncetion with Hermann Weyl's inversion problem namely "how to hear
the shape of a drum?" If one substitutes drum by the full-fledged QFT
containing globally charged fields and the perceived sound by the observable
charge neutral fields, the existence of the superselection theory stands for
the reconstruction of the full theory (containing all charges) from the
observable "shadow". The existence and uniqueness of this inverse problem has
given a significant insight into the inner workings of QFT; in particular it
demystified the origin of inner symmetries, a concept which started in the 30s
with Heisenberg's isospin. Morally something like this should also hold for
local (Maxwell) charges, if one allows fro the possibility that certain
charges cannot be registered and probably not even produced in collisions of
neutral particles (see section 7). However investigations in this direction
did not lead to anything tangible. We view our approach as yet another attempt
in this direction.

The violation of the Haag duality for the above doubly connected $\mathcal{T}$
has a quite different physical message. First it can be detected already in
the Wigner setting, so it has nothing to do with superselection sectors and
vacuum polarization. To appreciate its message, it is indicative to imagine
that the field strength can be derived from a pointlike potential $A_{\mu}(x)$
which is the standard starting point of the indefinite metric gauge setting$.$
In that case the fluxes will be supported inside the spacelike separated
$\mathcal{T}$. The vanishing of the resulting expression is in in flagrant
violation of the above calculations. This shows that the the standard
indefinite metric gauge formalism is unreliable. As we pointed out before it
is not wrong, but one has to carefully distinguish situations where it can be
applied from those where it leads to incorrect conclusions\footnote{The point
here is that in gauge theoretic calculations one computes numbers and is
normally one is not interested in localization. On the other hand the
commitment to gauge invariant results is mainly a lip-service, physical
correlations are easily characterized in terms of BRST invariance, but the
computation of charge neutral composites is quite a different story.}. For the
present purpose it is the \textit{strongest support for the introduction of
stringlike potentials} in the absence of interactions which would not create
any contradiction in the above calculation.

Both, the charge superselection problem and the problem of multi-connectedness
are intimately related to the way in which models of local quantum physics
realizes the localization principle. The Aharonov Bohm effect is perhaps the
most direct and simple illustration since it does not require composite fields
and vacuum polarization.

To generalize this subtle violation of Haag duality to arbitrary $(m=0,s>1)$
representations one needs a more adequate modular setting than the pointlike
covariant field strength formalism used in \cite{LRT}; the latter becomes
increasingly complicated with the increasing number of tensor/spinor indices.
A structural method which avoids the use of covariant field coordinatizations
and which is a wave funtion preform of the net of localized algebras in the
LQP formulation of QFT consists in the use is the use of the Wigner
representation theory of the Poincare group. In order to loose the reader, I
will try to use one paragraph to present at least some of the physical content
of that impressive theory; for its mathematical backup see \cite{Summers}

It has been known for a long time that the algebraic strucure underlying free
fields allows a functorial interpretation in which operator subalgebras of the
global algebra $B(H)$ are the functorial images of subspaces of the Wigner
wave function spaces ("second quantization"\footnote{Not to be confused with
quantization; to quote a famous saying by Ed Nelson: "quantization is an art,
but second quantization is a functor".}), in particular the spacetime
localized algebras are the images of localized subspaces. Since localized
subalgebras in QFT $\mathcal{A(O})$ are known to act cyclic and separating on
the vacuum (the Reeh-Schlieder property), the conditions for the validity of
the Tomita-Takesaki modular theory are fulfilled.

The start for constructing such subspaces K is always the localization in
wedge regions W, since in that case the modular objects associated with the
Tomita S-operator $S=J\Delta^{\frac{1}{2}}$ are part of Wigner's
representation theory, namely the antiunitary $J$ and modular unitary
$\Delta^{it}$ coalesce with the spacetime reflection on the edge of the wedge
and with the wedge-preserving boost respectively \cite{Haag}. The real
localization subspace $K(W)$ is just the closed $+1$ eigenspace of S and the
associated dense complex standard space is its complexification%
\begin{align}
&  K(W)=\overline{\{\varphi|S\varphi=\varphi\}}\label{S}\\
&  K(W)+iK(W)~is~standard\nonumber
\end{align}
where standardness (K and iK have trivial intersection and taken together form
a dense set) of spaces and operator algebras (intimately related to the
Reeh-Schlieder property \cite{Haag}) is one of the most important concepts of
modular localization theory.

The modular objects for subwedge regions are determined by representing them
in terms of intersections of wedges and showing the standardness of the
associated subspaces. This is fairly easy for noncompact regions as spacelike
cones $\mathcal{C}$, whose core is a semiinfinite spacelike string with an
arbitrary small opening angle. Evidently they can be represented by
intersections of wedges with a shared origin which then becomes the apex of
the spacelike cone $\mathcal{C}.$ In this case the standardness follows from
the energy positivity \cite{BGL} i.e. it is shared by all three Wigner
representation. All 3 families have stringlike wave-function valued generators
$\Psi(x,e),$ but only in the infinite spin case there are no better localized generators.

A systematic investigations along these purely modular lines would start with
showing the standardness of compact \ double cone $\mathcal{D}$ localized
subspaces $K(\mathcal{D})$ avoiding the use of covariant wavefunctions/fields
and relying entirely on the Wigner representation theory. The double cones are
the causally closed regions in terms of which the setting of algebraic QFT
(spacetime-indexed nets) is defined. In that case the origin of the wedges
cannot be fixed (to the origin of the Minkowski spacetime coordinatization)
but have to be passed around a circle on the two-dimensional spatial boundary
if a symmetrically chosen (around the coordinate origin) double cone. From
concrete calculations with pointlike generating fields we know that generating
wave functions $\Psi^{(A,\dot{B})}(x)$ obtained from covariantization of the
Wigner wave function do generate standard compactly localized subspaces. But
only an abstract version of this proof will reveal the importance of the
nature of the little group and its impact on the localization problem, i.e.
why unitary representations of compact- and finite dimensional representations
of noncompact little groups are compactly localizable, whereas the infinite
spin representation is not ($K(\mathcal{D})=0$).

The modular localization of states is much weaker than that of local algebras.
Whereas for positive energy states on which the Poincar\'{e} group is
unitarily represented the modular localization leads can be expressed in terms
of an antilinear operator which for wedge localization is of the form
$S=J_{0}\Delta$

Having prepared oneself in this way, there remains the structural
understanding of the generalized Aharonov-Bohm effect namely for multiply
connected regions there exist observables which, although they commute with
the field strength in the causal disjoint, are not expressible in terms of
field strength inside the multiply connected spacetime region. Since the field
strength determine the global properties, the generalized A-B effect is
another manifestation of the holistic aspects of QFT in this case one which
distinguishes massive from massless representation.

Although the QFT A-B effect has only been established for (m=0, s=1), the role
of the little group in localization leaves little doubt that there exists a
generalization for s%
$>$%
1. Invoking a metaphoric principle namely that nature may have only few
principles but an enormous variety of different manifestations, one is
inclined to speculate that the increasing number of potentials with increasing
s is associated with higher than double connectivity generalizations of the
QFT A-B effect (and not only with an increase of the number of A-B operators
in a geometric situation of n separated $\mathcal{T}$ regions. This would shed
light into the dark corners of higher spin quantum matter which has been
closed to gauge theory inspired ideas. Finally there is the question of the
action of the modular group in massless theories, especially in cases where
one expects this to be geometric, as in the case of $\mathcal{T}$. The answer
to any of these questions would require more mathematics and would lead too
far away from the spirits of this paper for which this section only serves to
illustrate that there are indication for the role of string-localized
potentials already in the free field theories. But the importance of this
unexplored suggests to return to it in a more specific future context.

The zero mass higher spin field strengths exhibits the above increase of scale
dimension of pointlike generated field strength with spin and therefore shows
a worsening of field strength associated short distance singularities. The
bosonic potentials, namely n=s string-localized generators with increasing
short distance dimensions fill the gap between $d_{sca}$=1 up to d=s where s+1
is the $d_{sca}$ of the lowest dimensional field strength (the lowest
dimension consistent with the second line in (\ref{line})). We already
emphasized that the localization in a indefinite metric setting has no
relation to the physical localization; this is the main message of the A-B
effect and the violation of Haag duality for QFT.

The same second line (\ref{line}) contains the considerably reduced number of
spinorial descriptions for zero mass and finite helicity, although in both
cases the number of pointlike generators which are linear in the Wigner
creation and annihilation operators \cite{MSY}.

By using the recourse of string-localized generators $\Psi^{(A,\dot{B})}(x,e)$
one can \textit{restore the full spinorial spectrum} for a given $s,$ i. e.
one can move from the second line to the first line in (\ref{line}) by
relaxing the localization. Even in the massive situation where pointlike
generators exist but have short distance singularities which increase with
spin. there may be good reasons (lowering of short distance dimension down to
$d_{sca}$=1) to use string-like generators. In all cases these generators are
covariant and "string-local"

The explicit verification of stringlike locality is cumbersome because there
are no simple x-space formulae for stringlike Pauli-Jordan functions. It is
easier to avoid manifest x-space localization formulas and work instead with
intrinsically defined modular localization subspaces. In fact the construction
of the singular stringlike generators are not based on any gauge theory
argument but rather a consequence of the availability of stringlike
intertwiners for all unitary positive energy representations of the Poincare
group. In the present setting the constraining equations (\ref{Lor}) have
nothing to do with a gauge condition but are rather a consequence of
constructing intertwiners which localize on a string $x+\mathbb{R}_{+}e$ which
is the next best possibility in cases where the compact localized subspaces
are empty and their pointlike generators nonexistent. The pointlike aspect of
the gauge formalism is only physically relevant in case of gauge invariant
operators i.e. the pointlike generated e-independent subalgebra coalesces with
the gauge invariant subalgebra. So the stringlike approach complements the
gauge invariant construction by incorporating the charged sector of QED with
its infraparticle aspects and hopefully also the nonlocal aspects of gluons
and quarks which are the key to their "invisibility".

Whereas free vectorpotentials have a harmless string localization since by
applying a differential operator one can get rid of the semiinfinite string
and return to the pointlike field strength, we will see in the next section
that the interaction furnishes the charge carrying operators with a much more
autonomous stringlike localization which cannot be removed by differential
operators and in fact is intrinsic to the concept of electric charge.

The noninteracting $(0,s=2)$ representation is usually described in terms of
pointlike field strength in form of a 4-degree tensor which has the same
permutation symmetries as the Riemann tensor (often referred to as the
\textit{linearized Riemann tensor}) with $d_{sca}=3$ whereas its string
localized covariant potential $g_{\mu\nu}(x,e)$ has the best possible
dimension $d_{sca}~g=1.$ By allowing string localized potential one can for
all $(m=0,s\geq1)$ representations avoid the increase in the dimensions with
growing spin in favor of $d_{sca}=1$ (independent of spin) stringlike
potentials from which one may return to the pointlike field strengths by
applying suitable differential operators. In the massive case there is no
reason for doing this from the point of localization rather the only physical
reason for using the string like counterparts for the pointlike fields is
their lower short distance dimensions; again the optimal value is $d_{sca}$=1
for all spins. Hence candidates for renormalizable interactions in the sense
of power counting exist for all spins.

In order to be able to continue with the standard pointlike perturbative
formalism one took recourse to the Gupta-Bleuler or BRST gauge formalism. At
the end one has to extract from the results of the pointlike indefinite metric
calculations the physical data i.e. perturbative expressions in a Hilbert space

In this respect there is a significant conceptual distinction between e.g.
classical ED and QED which is masked by the joint use of the same terminology
"gauge". Whereas in classical theory the use of the gauge potential simplifies
calculations and leads to interesting connections with the geometry, in
particular with the mathematics of fibre-bundles, the quasiclassical treatment
of quantum mechanics in a classical external electromagnetic environment leads
to the Aharonov-Bohm effect which is usually considered as the physical
manifestation of the vectorpotential.

Finally in the quantum field theoretic setting of QED it becomes indispensable
since without the minimal coupling of quantum matter to the potential it would
be impossible to formulate QED. In this case the pragmatic meaning of the
terminology "gauge principle" stands for the continued use of the standard
pointlike field formalism of QFT within an indefinite metric setting and the
return via gauge invariance to a restricted Hilbert space setting in which the
formal pointlike localization is the same as the physical localization. The
string-localized approach is strictly speaking not a gauge theoretic
formulation in this sense. But neither is the closely related "axial gauge"
formulation since the axial potential already lives in a Hilbert space and
hence its localization is already physical. Although the clash between
pointlike localization and Hilbert space representation continues to hold for
the \ "potentials" of all $(m=0,s\geq1)$ representations, the analog of gauge
theory does not exist or is not known. It seems that in those cases there is
no "fake" pointlike formalism which can be corrected by a "gauge principle"
which then selects the genuinely pointlike observables from the fake objects;
in those cases one has to face the issue of string-localized fields right from
the beginning.

The next interesting case beyond $s=1$\footnote{We omit spinor fields, as the
zero mass Rarita-Schwinger representation (m=0,s=3/2)$.$} is ($m=0,s=2$); in
that case the "field strength" is a fourth degree tensor which has the
symmetry properties of the Riemann tensor; in fact it is often referred to as
the linearized Riemann tensor. In this case the string-localized potential is
of the form $g_{\mu\nu}(x,e)$ i.e. resembles the metric tensor of general
relativity. The consequences of this localization for a reformulation of gauge
theory will be mentioned in a separate subsection.

\section{String-localization of charged states in QED and Schwinger-Higgs
screening}

In this section some consequences of working with physical\footnote{Here we do
not distinguish between "physical" and "operator in a Hilbert space" i.e.
"unphysical" refers to an object in an indefinite metric space. Of course they
maybe good reasons to further restrict this terminology within a Hilbert space
setting in a more contextual way.} i.e. string-localized vector potentials in
perturbatively interacting models will be considered. Whereas all charge
neutral objects in QED are pointlike generated, this cannot be true for
physical charge-carrying operators. From the previous sections we know that
the best noncompact localized charged generators are semiinfinite spacelike
strings which, as a result of their simultaneous fluctuations in the Minkowski
spacetime $x$ and the spacelike direction $e$ (3-dim. de Sitter) have improved
short distance behavior in $x$, namely there always exists a potential with
$d_{sca}=1$ which is the best short distance behavior which the Hilbert space
positivity allows for a vectorpotential associated with a field strength of
scaling dimension $d_{sca}=2$. The prize to pay for part of the field strength
fluctuations having gone into the fluctuation of the string direction $e$ is
the appearance of infrared divergences which require the distribution
theoretical treatment of the variable $e;$ this problem must be taken care of
with special care in perturbative calculations.

In the previous section we learned that the full covariance spectrum
(\ref{line}) for zero mass finite helicity representation can be regained by
admitting stringlike fields. The pointlike \textit{field strength}
\footnote{We use this terminology in a generalized sense; all the pointlike
generators (the only ones considered in \cite{Wein}) are called field strength
(generalizing the $F_{\mu\nu}$) whereas the remaining string-localized
generators are named potentials.} is then connected with the
\textit{stringlike potentials }by covariant differential operators. We have
presented structural arguments in favor of using stringlike potentials over
field strength even in the absence of interactions when stokes argument is
used to rewrite a quantum magnetic flux integral over a surface into an
integral over its boundary. However the most forceful argument is that for
each spin $s\geq1$ there exists always a potential of lowest possible
dimension namely $d_{sca}(\Psi^{(\frac{s}{2},\frac{\dot{s}}{2})}(x,e))=1$
which is the power-counting prerequisite for constructing renormalizable interactions.

This holds also in the massive case where the covariance for pointlike fields
covers the whole spinorial spectrum (\ref{line}). Whereas the pointlike fields
have an $d_{sca}\geq1$ which increases with $s,$ there also exist stringlike
fields with $d_{sca}=1$ for any $s$. The simplest example would be a massive
pointlike vector field $A_{\mu}(x)$ with short distance dimension $d_{ssd}=2$
and a stringlike potential $A_{\mu}(x,e)$ with dimension $d_{ssd}=1.$ It is
only the stringlike potential which has a massless limit.

In this case there is no representation theoretical reason to introduce them
(no clash of localization and positivity), rather the only reason for doing
this is to meet the power-counting preconditions for renormalizability.
Whereas with pointlike fields the power-counting short distance restriction of
maximal $d_{sd}(interaction)=4$ only allows a finite number of low spin
models, the stringlike situation increases this number to infinite, since now
power-counting renormalizable interactions with string-localized potentials of
arbitrary high spin exist. For example for the string-localized s=2 symmetric
tensor potential $g_{\mu\nu}(x,e)$ there exist interactions which obey the
power-counting condition, but this of course does not mean that specific
interesting models with pointlike observables as the Einstein-Hilbert action,
are among this larger class of renormalizable candidates. Instead of searching
for a gauge principle which singles out pointlike generated observables (e.g.
the Riemann tensor $R_{\mu\nu\kappa\lambda}$), the problem one faces is to
understand the relation between a coupling dependent law for the change of
potentials under the change of string direction $e\rightarrow e^{\prime}$ and
the form of the pointlike composites.

It was already mentioned that the string-localization has hardly any physical
consequences for photons, since even in the presence of interactions the
content of the calculated theory can be fully described in terms of linearly
related pointlike field strengths. Even the scattering theory of photons in
the charge zero sector has no infrared problems. However the
interaction-induced string-localization of the charged field which is
transferred from the vectorpotentials\footnote{Localization of the free
fields, in terms of which the interaction is defined in the perturbative
setting, is not individually preserved in the presence of interactions; the
would be charged fields are not immune against delocalization from
interactions with stringlike vectorpotentials.} is a more serious matter; it
is inexorably connected with the electric charge, and there is no linear
operation nor any other manipulation which turns the noncompact localization
of charged quantum matter into compact localization. The argument (\ref{Ha})
based on the use of the quantum adaptation of Gauss's law shows that the
noncompact (at best stringlike) localization nature of generating Maxwell
charge-carrying fields is not limited to perturbation theory.

Its most dramatic observable manifestation occurs in scattering of charged
particles. As mentioned before, the infrared peculiarities of scattering of
electrically charged particles were first noted by Bloch and Nordsiek, but no
connection was made with the string-localization which was
suggested\footnote{Localization properties in terms of gauge dependent fields
are not necessarily physical.} at the same time by the formula (\ref{DJM})
from gauge theory. One reason is certainly that the standard perturbative
gauge formalism (which existed in its non-covariant unrenormalized form since
the time of the B-N paper) was not capable to address the construction of
string-localized physical fields. This is particularly evident in renormalized
perturbation theory which initially seemed to require just an adaptation of
scattering theory \cite{YFS}, but whose long term consequences, namely a
radical change of one-particle states and the spontaneous breaking of Lorentz
invariance, were much more dramatic.

These phenomena were incompletely described in the standard perturbation
theory of the gauge setting which had no convincing practicable way to extend
the requirement of gauge invariance to the charged sectors. In particular the
observable part of the scattering formalism culminated in a calculational
momentum space recipe for inclusive cross sections; it was not derived in a
spacetime setting as the LSZ scattering formalism for interactions of
pointlike fields. The spacetime setting in a theory as QFT, for which
everything must be reduced to its localization principles, is much more
important than in QM where stationary scattering formulations compete with
time-dependent ones. As mentioned before Coulomb scattering in QM can be
incorporated into any formulation of scattering theory by extracting a
diverging phase factor which results from the long range. Noncompact
string-localization is a more violent change from pointlike generated QFT than
long- versus short range quantum mechanical interactions.

Perturbative scattering (on-shell) processes represented by graphs which do
not contain inner photon lines turn out to be independent of the string
direction $e$ i.e. they appear as if they would come from a pointlike
interaction\footnote{The time-ordered correlation functions, of which they are
the on-shell restriction, are however string-dependent.}. This includes the
lowest order M\o ller- and Bhaba scattering. The mechanism consists in the
application of the momentum space field equation to the u,v spinor wave
functions so that from (\ref{tp}) only the $g_{\mu\nu}$ term in the photon
propagator survives. The terms involving photon lines attached to external
charge lines do however depend on the string directions, and if the scattering
amplitudes would exist (they are infrared divergent), they would be
e-independent. The on-shell infrared divergence and the e-dependence are two
sides of the same coin. One expects the photon inclusive cross section to be
finite and e-dependent (at least in the low energy domain). By using the
additional resource of e-smearing one expects for the first time the possible
formulation of a large time convergence aiming directly at inclusive cross sections.

In the sequel some remarks on the perturbative use of stringlike
vectorpotentials for scalar QED are presented which is formally defined in
terms of the interaction density\footnote{The integral over the interaction
density is formally e-independent.}%
\begin{equation}
g\varphi(x)^{\ast}(\partial_{\mu}\varphi(x))A^{\mu}(x,e)-g(\partial_{\mu
}\varphi(x)^{\ast})\varphi(x)A^{\mu}(x,e) \label{coup}%
\end{equation}
It is also the simplest interaction which permits to explain the Higgs
mechanism as a QED charge-screening. The use of string-localized
vectorpotentials as compared to the standard gauge formalism deflects the
formal problems of extracting quantum data from an unphysical indefinite
metric setting to the ambitious problem of extending perturbation theory to
the realm of string-localized fields. This is not the place to enter a
presentation of (yet incomplete) results of a string-extended Epstein-Glaser
approach. Fortunately this is not necessary if one only wants to raise
awareness about some differences to the standard gauge approach.

It has been known for a long time that the lowest nontrivial order for the
Kallen-Lehmann spectral function can be calculated without the full
renormalization technology of defining time-ordered functions. With the field
equation
\begin{equation}
(\partial^{\mu}\partial_{\mu}+m^{2})\varphi(x)=gA_{\mu}(x,e)\partial^{\mu
}\varphi(x)
\end{equation}
the two-point function of the right hand side in lowest order is of the form
of a product of two Wightman-functions namely the point-localized
$\left\langle \varphi(x)\varphi^{\ast}(y)\right\rangle =i\Delta^{(+)}(x-y)$
and that of the string-localized vectorpotential (\ref{tp})%
\begin{equation}
\left\langle A^{\mu}(x,e)A^{\nu}(x^{\prime},e^{\prime})\right\rangle
\left\langle \partial_{\mu}\varphi(x)\partial_{\nu}\varphi^{\ast}(x^{\prime
})\right\rangle
\end{equation}
leading to the two-point function in lowest (second) order
\begin{equation}
(\partial_{x}^{2}+m^{2})(\partial_{x^{\prime}}^{2}+m^{2})\left\langle
\varphi(x)\varphi^{\ast}(x^{\prime})\right\rangle _{e,e^{\prime}}^{(2)}\sim
g^{2}\left\langle A^{\mu}(x,e)A^{\nu}(x^{\prime},e^{\prime})\right\rangle
\left\langle \partial_{\mu}\varphi(x)\partial_{\nu}\varphi^{\ast}(x^{\prime
})\right\rangle
\end{equation}
which is manifestly $e$-dependent in a way which cannot be removed by linear
operations as in passing from potentials to field strength. One can simplify
the $e$ dependence by choosing collinear strings $e=e^{\prime},$ but the
vectorpotential propagator develops an infrared singularity and in general
such coincidence limits (composites in d=2+1 de Sitter space) have to be
handled with care (although these objects are always distributions in the
string direction i.e. can be smeared with localizing testfunctions in de
Sitter space); just as the problem of defining interacting composites of
pointlike fields through coincidence limits. The infrared divergence can be
studied in momentum space; a more precise method uses the mathematics of wave
front sets\footnote{Technical details as renormalization, which are necessary
to explore these unexplored regions, will be deferred to seperate work.}. This
simple perturbative argument works for the second order two-point function,
the higher orders cannot be expressed in terms of products of Wightman
function but require time ordering and the Epstein-Glaser iteration.

Not all functions of the matter field $\varphi$ are $e$-dependent; charge
neutral composites, as e.g. normal products $N(\varphi\varphi^{\ast}$)$(x)$ or
the charge density are $e$-independent. On a formal level this can be seen
from the graphical representation since a change of the string direction
$e\rightarrow e^{\prime}$ (\ref{change}) corresponds to an abelian gauge
transformation. The divergence form of the change of localization directions
together with the current-vectorpotential form of the interaction reduces the
$e$-dependence of graphs to \textit{vectorpotentials propagators attached to
external charged} lines while all $e$-dependence in loops cancels by partial
integration and current conservation. This is in complete analogy to the
standard statement that the violation of gauge invariance and the cause of
on-shell infrared divergencies on charged lines result from precisely those
external charge graphs; external string-localized vectorpotential lines cause
no problems since they loose their $e$-dependence upon \ differentiation. A
\textit{neutral external composite} as $\varphi\varphi^{\ast}$ on the other
hand does not generate an external charge line; again the gauge invariance
argument parallels the statement that such an external vertex does not
contribute to the string-localization.

Hence both the gauge invariance in the pointlike indefinite metric formulation
and the $e$-independence in the string-like potential formulation both lead to
pointlike localized subtheories\footnote{Note however that the spacetime
interpretation of the $e$ is not imposed. The proponents of the axial gauge
could have seen in in the free two-pointfunction of vectorpotentials and in
all charge correlators if they would have looked at the commutators inside
their perturbative correlation functions. The axial "gauge" is not a gauge in
the usual understanding of this terminology.}. But whereas the embedding
theory (Gupta-Bleuler, BRST) in the first case is unphysical\footnote{The
pointlike localization in an indefinite metric description is a fake. Its
technical use is that pointlike interactions, whether in Hilbert space or in a
indefinite matric setting, come with a well known formalism. The gauge
invariant correlation define (via the KMS construction) a new Hilbert space
which coalesces with the subspace obtained by application of the pointlike
generated subalgebra of the physical string-like formulation to the vacuum.},
the string-like approach uses Hilbert space formulations throughout. The
pointlike localization in an indefinite metric description is a fake. Its
technical advantage is that pointlike interactions, whether in Hilbert space
or in a indefinite metric setting, are treatable with the same well known
formalism. The gauge invariant correlation define (via the GNS construction) a
new Hilbert space which coalesces with the subspace obtained by application of
the pointlike generated subalgebra of the physical string-like formulation to
the vacuum.

But whereas the noncompact localized charge-carrying fields are objects of a
physical theory it has not been possible to construct physical charged
operators through Gupta-Bleuler of BRST cohomological descent. The difficulty
here is that one has to construct non-local invariants under the nonlinear
acting formal BRST symmetry. So the simplicity of the gauge formalism has to
be paid for when it comes to the construction of genuinely nonlocal objects as
charged fields.

This leaves the globally charge neutral \textit{bilocals} in the visor. Their
description is expected to be given in terms of formal \textit{bilocals which
have a stringlike "gauge bridge"} linking the end points of the formal
bilocals (\ref{br}). In contrast to the string-localized single operators it
is difficult to construct them in perturbation theory starting from
string-localized free fields, they are too far removed from the form of the
interaction (see also next section). In order to understand the relation
between such neutral bilocals and infraparticles one should notice that in
order to approximate a scattering situations, the "gauge-bridge" bilocals will
have to be taken to the limiting situation of an infinite separation distance,
so that the problem of the infinite stringlike localization cannot be avoided
since it returns in the scattering situation. The only new aspect of the
proposed approach based on string-localized potentials which requires
attention is that the dependence on the individual string directions $e$ is
distributional i.e. must be controlled by (de Sitter) test function smearing
and moreover that composite limits for coalescing $e^{\prime}s$ can be defined.

Finally there is the problem of Schwinger-Higgs mechanism in terms of string
localization. The standard recipe starts from scalar QED which has 3
parameters (mass of charged field, electromagnetic coupling and quadrilinear
selfcoupling required by renormalization theory). The QED model is then
modified by Schwinger-Higgs screening in such a way that the Maxwell structure
remains and the total number of degrees of freedom are preserved. The standard
way to do this is to introduce an additional parameter via the
vacuumexpectation value of the alias charged field and allow only
manipulations which do not alter the degrees of freedom. We follow Steinmann
\cite{Stei}, who finds that the screened version consists of a selfcoupled
real field $R$ of mass $M$ coupled to a vectormeson $A^{\mu}$ of mass $m$ with
the following interaction%
\begin{align}
L_{int}  &  =gmA^{\mu}A_{\mu}R-\frac{gM^{2}}{2m}R^{3}+\frac{1}{2}g^{2}A_{\mu
}A^{\mu}R^{2}-\frac{g^{2}M^{2}}{8m^{2}}R^{4}\label{scre}\\
\Psi &  =R+\frac{g}{2m}R^{2}%
\end{align}
The formula in the second line is obtained by applying the prescription
$\varphi\rightarrow\left\langle \varphi\right\rangle +R+iI$ to the complex
field within the neutral (and therefore point-local) composite $\varphi
\varphi^{\ast}$ and subsequently formally eliminating the $I$ field by a gauge
transformation. The result is the above interaction where $A_{\mu}$ and $R$
are now massive fields. Since the field $\Psi$ is the image of a pointlike
$\varphi\varphi^{\ast}$ under the Higgs prescription, the real matter field
$\Psi$ is point-local.

The important point which formalizes the meaning of "screening" is that the
algebraic Maxwell structure as well as the degrees of freedom remain
preserved\footnote{The degrees of freedom of the real massive field I(x) went
into the conversion of a photon into a massive vectormeson.} even though the
interaction in terms of the new fields $R$ and the massive vectorpotential
$A_{\mu}(x,e)$ breaks the charge symmetry (by "screening" i.e. trivializing
the charge, see below) and the even-odd symmetry $R\rightarrow-R$ of the
remaining $R$-interaction. It is this discrete symmetry breaking which renders
the even-odd selection rule ineffective and by preventing that $R$ can have a
different localization from $R^{2}$ the pointlike localization of the
quadratic terms is transferred to the linear $R.$ The stringlike $d_{sca}=1$
massive vectormeson $A_{\mu}(x,e)$ played the important technical role in the
renormalizability of the theory but is not needed to describe the constructed
theory in terms of generating fields: a pointlike $F_{\mu\nu}$ (or an
associated pointlike $A_{\mu}(x)$) and a pointlike $R(x).$

Hence in the present context the string-localized potentials, as well as the
BRST formalism, behave as a "catalyzer" which makes a theory amenable to
renormalization. The former have the additional advantage over the latter that
the Hilbert space is present throughout the calculation.

One has to be careful in order not to confuse computational recipes with
physical concepts. Nonvanishing vacuum expectations (one-point functions) are
part of a recipe and should not be directly physically interpreted, rather one
should look at the intrinsic observable consequences\footnote{These are
properties which can be recovered from the observables of the model i.e. they
do not depend on the particular method of construction.} before doing the
physical mooring. The same vacuum expectation trick applied to the Goldstone
model of spontaneous symmetry breaking has totally different consequences from
its application in the Higgs-Kibble (Brout-Englert, Guralnik-Hagen) symmetry breaking.

In the case of spontaneous symmetry breaking (Goldstone) the \textit{charge
associated with the conserved current diverges} as a result of the presence of
a zero mass Boson which couples to this current. On the other hand in the
Schwinger-Higgs screening situation \textit{the charge of the conserved
current vanishes} (i.e. is completely screened) and hence there are no charged
objects which would have to obey a charge symmetry with the result that the
lack of charge resulting from a screened Maxwell charge looks like a symmetry
breaking.%
\begin{align*}
&  Q_{R,\Delta R}=\int d^{3}xj_{0}(x)f_{R,\Delta R}(x),~~f_{R,\Delta R}(x)=%
\begin{array}
[c]{c}%
1~for~\left\vert \mathbf{x}\right\vert <R\\
0~~for~\left\vert \mathbf{x}\right\vert \geq R+\Delta R
\end{array}
\\
&  \lim_{R\rightarrow\infty}Q_{R,\Delta R}^{spon}\left\vert 0\right\rangle
=\infty,~m_{Goldst}=0;~~\lim_{R\rightarrow\infty}Q_{R,\Delta R}^{screen}%
\psi=0,\text{ }all~m>0
\end{align*}

That the recipe for both uses a shift in field space by a constant does not
mean that the physical content is related. The result of screening is the
vanishing of a Maxwell charge which (as a result of the charge superselection)
allows a copious production of the remaining $R$-matter. Successful recipes
are often placeholders for problems whose better understanding needs
additional conceptual considerations. In both cases one can easily see that
the incriminated one-point vacuum expectation has no intrinsic physical
meaning, i.e. there is nothing in the intrinsic properties of the observables
of the two theories which reveals that a nonvanishing one-point function was
used in the recipe for its construction. For a detailed discussion of these
issues see \cite{Swieca}.

The premature interpretation in terms of objects which appear in calculational
recipes tends to lead to mystifications in particle theory; in the present
context the screened charged particle has been called the "God particle". As
mentioned before the Schwinger-Higgs screening is analog to the quantum
mechanical Debeye screening in which the elementary Coulomb interaction passes
to the screened large distance effective interaction which has the form of a
short range Yukawa potential. The Schwinger-Higgs screening does not work
(against the original idea of Schwinger) directly with spinor- instead of
scalar matter. If one enriches the above model by starting from QED which
contains in addition to the charged scalar fields also charged Dirac spinors
then the screening mechanism takes place as above via the scalar field which
leads to a loss (screening, bleaching) of the Maxwell charge while the usual
charge superselection property of complex Dirac fields remains unaffected.

The Schwinger-Higgs mechanism has also a scalar field multiplet generalization
to Yang-Mills models; in this case the resulting multicomponent pointlike
localized massive model is much easier to comprehend than its "charged"
string-localized origin. As the result of screening there is no unsolved
confinement/invisibility problem resulting from nonabelian string-localization.

The Schwinger Higgs screening suggests an important general idea about
renormalizable interactions involving massive $s\geq1$ fields, namely that
formal power-counting renormalizability ($d_{sdd}=1$) is not enough. For
example a pure Yang-Mills interaction with massive gluons (without an
accompanying massive real scalar multiplet) could be an incorrect idea because
the string-localization of the Hilbert space compatible gluons could spread
all over spacetime or there may exist other reasons why the suspicion that
such theories are not viable may be correct. Such a situation would than be
taken as an indication that a higher spin massive theory would always need
associated lower spin massive particles in order to be localizable; in the s=1
case this would be the s=0 particle resulting from Schwinger-Higgs screening.
Before one tries to understand such a structural mechanism which requires the
presence of localizing lower spin particles it would be interesting to see
whether these new ideas allow any renormalizable s=$\frac{3}{2}$
(Rarita-Schwinger) theories. Even though there may be many formal
power-counting renormalizable massive $s\geq1$ interactions only a few are
expected to be pointlike localized.

It is interesting to mention some mathematical theorems which support the
connection between localization and mass spectrum. The support for placing
more emphasis on localization in trying to conquer the unknown corners of the
standard model comes also from mathematical physics. According to Swieca's
theorem\footnote{Actually Swieca does not use locality directly but rather
through its related formfactor analyticity which is different for
string-localized (Maxwell) charged particles (less analyticity) from neutal
massive sreened particles.} \cite{screening}\cite{Swieca} one expects that the
screened realization of the Maxwellian structure is local i.e. the process of
screening is one of reverting from the electromagnetic string-localization
back to point locality together with passing from a gap-less situation to one
with a mass gap. Last not least the charge screening leads to a Maxwell
current with a vanishing charge\footnote{Swieca does not directly argue in
terms of localization but rather uses the closely related analyticity
properties of formfactors.} and the ensuing copious production of alias
charged particles. The loss of the charge superselection rule in the above
formulas (\ref{scre}) is quite extreme, in fact even the $R\leftrightarrow-R$
selection rule has been broken (\ref{scre}) in the above Schwinger-Higgs
screening phase associated with scalar QED. The general idea for constructing
renormalizable couplings of massive higher spin potentials interacting with
themselves or with normal s=0,1/2 matter cannot rely on a Schwinger-Higgs
screening picture because without having a pointlike charge neutral subalgebra
for zero mass potentials as in QED, which is the starting point of gauge
theory, there is no screening metaphor which could preselect those couplings
which have a chance of leading to a fully pointlike localized theory, even
though renormalizability demands to treat all $s\geq1$ as stringlike objects
with $d_{sca}$=1. Of course at the end of the day one has to be able to find
the renormalizabe models which maintain locality of observables either in the
zero mass setting as (charge-neutral) subalgebras (QED,Yang-Mills) or the
massive theories obtained from the former with the help of the screening idea.
gauge theory is a crutch whose magic power is limited to s=1, for s$>1$ it
lost its power and one has to approach the localization problem directly.

The existence of a gauge theory counterpart, namely the generalization of the
BRST indefinite metric formalism to higher spins, is unknown. So it seems that
with higher spin one is running out of tricks, hence one cannot avoid confront
the localization problem of separating theories involving string-localized
potentials which have pointlike generated subalgebras from those which are
totally nonlocal and therefore unphysical. This opens a new chapter in
renormalization theory and its presentation would, even with more results than
are presently available, go much beyond what was intended under the modest
title of this paper.

An understanding of the Schwinger-Higgs screening prescription in terms of
localization properties should also eliminate a very unpleasant previously
mentioned problem which forces one to pass in a nonrigorous way between the
renormalizable gauge (were the perturbative computations take place) and the
"unitary gauge" which is used for the physical interpretation. The relation
between the two remains somewhat metaphoric.

In contradistinction to theories with string-localized electric charge
carrying infraparticles and the Schwinger-Higgs screening mechanism one can
study models with a discrete gauge groups on the lattice. In the case of a
$\mathbb{Z}_{2}$ gauge model the result is a rich phase diagram \cite{Fr-Ma}
in which also a phase is realized in which massive excitations (Higgs) exist
jointly with free string-localized $\mathbb{Z}_{2}$-charges. The authors
interprete this as a realization of massive strings in the sense of
\cite{B-F2}. The problem of the mathematical control of continuum limits
continues unabated.

The screened interaction between a string-localized massive vectorpotential
and a real field (\ref{scre}) remains pointlike because the string
localization of the massive vectorpotential only serves to get below the power
counting limit but does not de-localize the real matter field; since the
pointlike field strength together with the real scalar field generate the
theory, the local generating property holds. In an approach based on
string-localization there is only one description which achieves its
renormalizability by string-localized potentials.

The BRST technology is highly developed, as a glance into the present
literature \cite{Du} shows. It certainly has its merits to work with a
renormalization formalism which starts directly with massive vectormesons
\cite{Du-S} instead of the metaphoric "photon fattened on the Higgs one point
function". It is hard to think how the BRST technology for the presentation of
the Schwinger-Higgs screening model which starts with a massive vectormeson in
\cite{Du} can be improved. For appreciating this work it is however not
necessary to elevate \ "quantum gauge symmetry" (which is used as a technical
trick to make the Schwinger-Higgs mechanism compliant with renormalizability
of massive s=1 fields) from a useful technical tool to the level of a new principle.

Besides the Schwinger-Higgs screening mechanism which leads to renormalizable
interactions of massive vectormesons with low spin matter, there is also the
possibility of renormalizable "massive QED" which in the old days \cite{Lo-S}
was treated within a (indefinite metric) gauge setting in order to lower the
short distance dimension of a massive vectormeson from $d_{sca}=2$ to $1$, and
in this way stay below the powercounting limit. Such a construction only works
in the abelian case; for nonabelian interactions the only way to describe
interacting massive vectormesons coupled to other massive s=0,1/2 quantum
matter is via Higgs scalars in their Schwinger-Higgs screening role. Whereas
the local Maxwell charge is screened, the global charges of the non-Higgs
complex matter fields are preserved. It seems that Schwingers original idea of
a screened phase of spinor QED cannot be realized, at least not outside the
two-dimensional Schwinger model (two-dimensional massless QED).

But the educated conjectures in this section should not create the impression
that the role of the Schwinger-Higgs screening in the renormalizability of
interactions involving selfcoupled massive vectormesons has been completely
clarified; if anything positive has been achieved, it is the demystification
of the metaphor of a spontaneous symmetry breaking through the
vacuumexpectation of a complex gauge dependent field and the tale of "God's
particle" which creates the masses of s=1/2 quantum matter. Actually part of
this demystification has already been achieved in \cite{Du}.

This leads to the interesting question whether, apart from the presence of the
Higgs particle (the real field as the remnant of the Schwinger-Higgs
screening), there could be an intrinsic difference in the structure of the
vectormeson. Such a difference could come from the fact that the screening
mechanism does not destroy the algebraic structure of the Maxwell equation,
whereas an interaction involving a massive vectormeson coming in the indicated
way from a S-H screening mechanism and interacting with spinorial matter
fields maintains the Maxwell structure. In the nonabelian case this problem
does not arise since apparently the Schwinger-Higgs screening mechanism is the
only way to reconcile renormalizability with localizability (or a return to
physics from an indefinite metric setting).

This raises the interesting question whether renormalizability and pointlike
locality of interactions with massive higher spin $s>1$ potentials is always
related to an associated zero mass problem via an analog of a screening
mechanism in which a lower spin field plays the analog of the Higgs field?

Whereas for interactions between spin one and lower spin fields the physical
mechanism behind the delocalization of matter (or rather its noncompact
re-localization) is to some degree understood, this is not the case for
interacting higher spin matter. Stringlike interactions enlarge the chance of
potentially renormalizable (passing the power counting test) theories, in fact
stringlike potentials with dimension $d_{sca}=1$ exist for any spin (hence
infinitely many) whereas the borderline for pointlike interaction is $s=1/2$
and with the help of the gauge setting $s=1$. Certain interactions, as the
Einstein-Hilbert equation of classical gravity probably remain outside the
power-counting limit even in the stringlike potential setting, but certain
polynomial selfinteractions between the $g_{\mu\nu}(x,e)$ with $dimg_{\mu\nu
}(x,e)=1$ may be renormalizable. The existence of free pointlike field
strength (in this case the linearized Riemann tensor) indicates that there may
be renormalizable interactions which lead to pointlike subalgebras, but the
presence of self-couplings modifies the transformation law under a change of
$e$ (\ref{change}) which now depends on the interaction as it is well-known
from the gauge theoretical formulation for Yang-Mills couplings.

One of course does not know whether QFT is capable to describe quantum
gravity, but if it does in a manner which is compatible with renormalized
perturbation theory, there will be no way to avoid string-localized
tensorpotentials even if the theory contains linear or nonlinear related
pointlike localized field strength. The trick of gauge theory, by which one
can extract pointlike localized generators without being required to construct
first the string-localized ones, is a resource which does not seem to exist
for higher spins, not even if one is willing to cope with unphysical ghosts in
intermediate steps. The most interesting interactions are of course the
selfinteractions between $(m=0,s>1).$ Here one runs into similar problems as
with Yang-Mills models (next section). The independence on $e^{\prime}s$ of
the local observables leads to nonlinear transformation laws which extend that
of free stringlike potentials and the non-existence of linear local
observables. Although saying this does not solve any such problem, the lack of
an extension of the gauge idea to higher spin makes one at least appreciative
of a new view based on localization.

There is one important case which we have left out, namely that of massless
Yang-Mills theories interaction with massive matter. This will be discussed in
the next section.

There are 2 different categories of delocalization: string-localization with
and without nontrivial pointlike-generated subalgebras. Generically the
coupling of string-localized fields leads to a theory with no local
observables. The models of physical interest are those which contain
$e~$independent subfields. For the case at hand the crucial relation is that
the change in the string direction can be written as a derivative as in
(\ref{change}).

\subsection{A note on massive strings}

Historically there is a close relation between charge screening, which
re-converts string-like generators to pointlike ones, and a powerful theorem
\cite{B-F2} which limits the localization of massive charge-carrying
generators on nets of observable algebras to be at worse stringlike (pointlike
is a special case). In particular there is no usage for generating "branes".
The proof uses the connection between smoothness, and analyticity with
localization \cite{B-F1}. Such massive strings lead to the same scattering
theory as pointlike fields in the situation of a mass gap, even the crossing
property of formfactors seems to be the same although more detailed on-shell
analytic properties may be different.

A perturbative realization of strings with a mass gap does not exist. But
perhaps this is the result of starting with pointlike free fields. As in the
Hilbert space formulation using string-like free potentials one should perhaps
start with string-localized massive fields which in addition would also widen
the possibilities for renormalizable interactions. Of course there is no
principle which says that a more fundamental theory as QFT has to jump over a
classical cane; there are good indications that Lagrangian (functional
integral) QFT only form a thin subset of all physically acceptable QFTs.

\section{A perturbative signal of "invisibility" and "confinement"?}

The reader may wonder why a concept as confinement, which for more than 4
decades has been with us and entered almost every discourse on strong
interactions appears in the title of this section in quotation marks. The
truth is that these issues are the least understood aspects of gauge theory,
and there are even reasons why this terminology may be questionable.

In QM a confinement into a spatial "cage" can be implemented by choosing a
confining potential; since the first models for quark-confinement were quantum
mechanical, this explains the origin of the terminology. However the
conceptual structure of QFT is radically different, and a spatial confinement
is not implementable in a setting of QT in which causal localization is the
main physical principle, one can at best attribute a metaphoric meaning:
confined from becoming on-shell. In fact it was (and still is) one of the
conjectures \cite{Swieca} in the 60s that compact localizability together with
a limitation on the phasespace degree of freedom\footnote{In the old days this
property was called "compactness", the modern somewhat stronger version is
called "nuclearity". The expression refers to the cardinality of states in a
phase space cell being equal to states in the range of a compact or
trace-class operator.} would result in "asymptotic completeness" i.e. the
property that every state of the theory can be written as a superposition of
(generically infinitely many) particle states. In contradistiction to QM this
cardinality per unit phase space volume is not finite, but a cardinality which
goes beyond nuclearity would also not be acceptable since it contradict the
causal propagation properties of Lagrangian quantization; more precisely it
leads to a violation of the requirement that there should not be more degrees
of freedom in the causal completion of a spacetime region $\mathcal{O}%
^{\prime\prime}$ than there were in the original region $\mathcal{O}$.
Theories in which this violation occurs have very weird properties: a kind of
poltergeist effect where additional degrees of freedom which were not present
in the initial region $\mathcal{O}$ are entering "sideways" into
$\mathcal{O}^{\prime\prime}.$ In case of manipulations outside the Lagrangian
formalism as holographic projections to lower dimensional QFTs, as the AdS-CFT
correspondence, this warning becomes very relevant \cite{causal}.

N-particle states, in such a LQP scenario are simply the stable n-fold counter
clicks in a coincidence/anticoincidence arrangement in which the spacetime
continuum has been cobbled with counters. Here stable means that a
counter-registered event of an n--fold excitation at large times does not
later turn any more into an m-state $m\neq n$ with changed velocities$.$ Only
in this case of asymptotic stability it makes sense to replace the word
"excitation" (of the vacuum) by "particle".

Part of this picture have been proven. It is true that a stable n-fold vacuum
excitation state is a tensor product state of n Wigner particles
\cite{Haag}\cite{Enss}. The only known physical counter example, the
electrically charged infraparticle states, which are obtained by applying a
smeared string-localized charged (Maxwellian charge) operator to the vacuum
and studying its asymptotic behavior, contains the mass shell component with
vanishing probability\footnote{The Hilbert space of QED does not contain a
Wigner particle with the mass of the electron.} and require a conceptually
different scattering treatment \cite{Porr}. "Infraparticles" are most suitably
described in terms of "weights" (a kind of singular state which cannot be
associated to a state vector), which are directly related to probabilities.
Fortunately the Wigner representation theory extends to weights.

No matter how much one compresses the momentum support against the lower value
$p^{2}=m^{2},$ one never arrives at a state which is not populated by
infinitely many photons; one can only control the value of the measuring
resolution $\Delta,$ but there will be always infinitely many photons with
energies below $\Delta$ which escape detection; if we refine our registering
precision i.e. $\Delta\rightarrow0,$ we do in the end not register anything
since the charged particles with sharp mass are not states but "weights"
\cite{Haag}. In other words there is a certain intrinsic lack of precision in
registering infraparticles in their states; assuming that the radiation of
photons is the only way by which we can measure the presence of charged
particles, the infraparticles with small $\Delta$ would not be perceived, in
agreement with the vanishing inclusive cross section for $\Delta\rightarrow0.$

But what happens in case of pure Yang Mills interaction of string-localized
gluons which live in a ghost-free physical space? Ignoring the fact that their
transformation law under changes of the string direction $e$ is more
complicated (interaction-dependent), one may point to the very nonlinear
structure of Yang-Mills potentials which have to play simultaneously the role
of the charged particles and the mediators of their interactions. But such a
vague metaphoric idea is no substitute for a realistic explanation. It is
however not unreasonable to think that there exist strings of different inner
tension. For abelian strings describing charged infraparticles the creation of
a pair of oppositely charged localized objects with a gauge bridge between
them, the probablility that (together with real photon clouds) they continue
separating to infinity in form of charged infraparticles is larger than that
for competing processes of changing into charge neutral outgoing particles is
a picture with a high credidility. The analog for nonabelian theories that to
the contary the probability of asymptotic separation is zero. If one adds that
the cause of the breaking of the gauge bridge and the conversion into ordinary
particles without a bridge is an increase in energy with the size of the
bridge we arrived at the standard metaphoric picture. But this cannot be the
end of the story because energy (except that of asymptotic particles) is not a
well-defined physical concept. In QFT which is solely built on the modular
localization principle each property or mechanism mut be reduced to this principle.

In view of the known perturbative (off-shell) infrared divergencies of all
correlation functions in nonabelian models, there can be no test of this idea
within the standard formulation of gauge theory. Passing to the description in
terms of string-localized potentials there exists at least a chance. The
reason is that a testfunction smearing in the $e^{\prime}s$ removes these
infrared divergences (= ultraviolet in d=1+2 de Sitter) controls these
divergencies and a limiting procedure similar to that in $x$ for composite
fields is expected lead to one $e$ for each external nonabelian quark or gluon
charge. This would be a perturbative off shell description in a bona fide
Hilbert space in which one expects to show that there is no nontrivial
scattering probability, not in ordinary scattering theory via a spacetime
limiting nor in terms of momentum space prescriptions for inclusive cross
sections. A construction of a bridged bilocal in terms of which one could
exemplify what happens when one enlarges the bridge separation would probably
remain out of reach as it is even in the abelian case of QED.

Another problem which should have a solution in this setting is the
construction of composite pointlike localized fields. \ Whereas in the
standard setting this cannot even be formulated, the absence of infrared
divergence and limiting procedures for coalescend $e^{\prime}s$ are expected
to play an important role in the classification of $e$-independent composites.
They would correspond to the composite gauge invariants as e.g. formally
$F^{2}$ whose construction in terms of infrared divergent correlations is an
ill-defined problem. Finally the problem of the occurrance of local gauge
groups would be even more demystified as it already became in the work of
Stora. This author showed that there is no necessity to prescribe a group in
addition to the number of Yang-Mills fields, rather the form of the
interaction follows from the number of selfinteracting string-localized
potentials and the existence of pointlike generated subalgebras; in the gauge
theoretic setting it would be a consequence of the consistency of the
nonabelian BRST formalism. Keeping in mind that the BRST symmetry is a formal
device, this recognition protects against wrong associations with physical
symmetries. All symmetries are at the end of the day related to localization
\cite{Haag}, but gauge "symmetries" are most directly related to the principle
of QFT than standard inner symmetries.

The closeness of string-localized potentials to the axial gauge should also
dispels the impression that the subject of the present paper is something very
speculative and distant. From a pragmatic viewpoint it is nothing else than
the attempt to make sense of the so called axial gauge by understanding the
origin of its apparent incurable infrared divergences and figuring out how
they arise from overlooked localization problems which cause fluctuating
string directions. There is certainly nothing more conservative than tracing
the infrared divergences to their origin and taming them by controlling the
fluctuation which cause them.

There is of course the still open problem of generalizing the Epstein Glaser
setting from pointlike to stringlike fields, a formalism which practitioners
of QFT hardly pay attention to since its impact on computation is
insignificant (renormalization as one goes along), but which in the new
context gains in importance. A spreading de-localization of quantum matter
which cannot be controlled in terms of string-like fields (but spreads with
the growing internal structure of Feynman graphs) would not be acceptable. It
is encouraging that the Epstein-Glaser iteration can be shown to work at least
if all $e~$are on a hyperplane \cite{Jens}.

In recent times methods taken from algebraic QFT have been applied to
perturbative gauge theory. Whereas in low spin $(s<1)$ QFT these methods lead
to a perturbative presentation of QFT, their application to gauge theory only
describe part of the theory. In QED the charged particles remain still outside
this formalism. Following Bogoliubov's generating time-ordered $S(f)$
functional formalism\footnote{This is a functional and not the S-matrix; its
connection with the latter requires special conditions (adiabatic limit).}, an
algebraic formulation in a compact region was defined and its limit to all
spacetime discussed (the algebraic adiabatic limit). In this approach one
avoids states and correlation functions and uses only operator-algebraic
structures; the box dependence is removed by a kind of "algebraic adiabatic
limit" \cite{Fred} and the role of the BRST construction becomes very clear
since the encounter with the infrared problems of correlation functions has
been shifted to the later task of constructing physical charges states. The
problem of states cannot be avoided if one wants to talk about localization,
Maxwell charges and their nonabelian counterparts; but the hope is that by
separating the algebraic structure from states, it takes on a more amenable
form. This approach should in principle permit the calculation of expectation
values of renormalized pointlike composites as $F^{2},$ since one expects that
their algebraic adiabatic limit exists. But such perturbative calculations do
not yet seem to exist.

Forgetting gauge theory for a moment, one may ask as a problem of principle
whether the de-localization in QFT can becomes so strong that an object cannot
be registered in a (always local) counter. Having no clues from interacting
models one may look at the only Wigner infinite spin representation
family\cite{MSY} for which there are no pointlike generators i.e. all
generators are stringlike. But whereas an interacting charged field applied to
the vacuum defines a state which can be decomposed with respect to the
Poincar\'{e} group into a continuum of pointlike generated components (this is
not a decomposition in the operator algebra), it seems to be impossible to
measure the "piece of the irreducible string state" which is localized inside
the compact (or at least quasi-compact \cite{Haag}) counter. A local
change\footnote{In order to be not bothered by vacuum polarization as a result
of sharp localization, it is customary to work with quasilocal counters
\cite{Haag}.} on an irreducible string state leads to problems with the
standard view about the measurement process. Even more problematic is the
creation of such an object as the result of scattering off ordinary particles.

The appearance of string-localized representations\footnote{In section 3 we
made a distinction between string-localized representations and zero mass
string-localized covariant potentials in pointlike generated (by "field
strengths") representations which do not exist as pointlike objects and whose
only mark on the representation is the A-B effect.} of the third Wigner class
(massless infinite spin) in gauge theories is not very plausible, since in a
perturbative setting the kind of irreducible representations of the
Poincar\'{e} group which appear in an interacting theory is believed to be
already decided by the zero order input. In other words, it is difficult to
conceive of an mechanism in perturbation theory whereby a free gluon potential
$A_{\mu}^{a}(x,e)$ (\ref{vector}) acting on the vacuum interacting with itself
passes to an object $A_{\mu}^{a}(x,e)_{i.s.}$ whose application to the vacuum
contains irreducible \textbf{i}nfinite \textbf{s}pin Wigner components
(\ref{inf}). However outside of perturbation theory this may not be true;
there exists presently no theorem which excludes the possibility that the
application of interacting gluons to the vacuum contains an irreducible
infinite spin representation component. Such components are inert, apart from
their coupling to gravity (since they carry nontrivial energy-momentum), and
therefore may better fit better to dark matter than to gluons/quarks.

Equally implausible is the presence of objects which only exist as composites
e.g. a Yang-Mills theory which consists only of $F^{2}$ without gluon degrees
of freedom. In the perturbative setting this would mean that the elementary
degrees of freedom, which in zero order perturbation theory were formally
present in the form of (point- or string-localized) free fields, are in fact a
fake in that the \textit{physical theory lives on a lesser number of degrees
of freedom,} which from the point of view of the original free fields that
went into the interaction density, would be considered as composites. The most
popular variant of this picture is that the degrees of freedom of the free
massive quark matter only served as a kind of "initial ignition" for getting
the perturbative interaction going, but that the Hilbert space in which the
interaction takes place has only composite pointlike local generators. But the
only known mechanism is the theory of superselected locally generated (only
pointlike generators) charges \cite{Haag} according to which one can recreate
the charge states from charge splitting and a "disposing the unwanted charge
behind the moon" argument \cite{Haag}. This argument has already its problem
with Maxwell charges and fails completely for gluons and quarks.

The most popular semi-phenomenological picture going into this direction is
that of globally color neutral \textit{bridged bilocals} quarks which are
supposed to break beyond a certain distance and pass to states consisting of
physical particles (hadronization). Here the bridge refers to the localization
of a connecting string. From a QFT-based conceptual (as opposed to QM) view
such a quark is an ontological chimera, since if the degrees of freedom are
not in the Hilbert space why should there exist a mock version up to a certain
distance. But as long as one keeps useful phenomenological ideas and the
mathematical conceptual content of a QFT apart, there is no problem and the
splitting string idea may serve as a useful placeholder of an unsolved problem.

Perhaps it is helpful to remind the reader of this very old construction. The
simplest illustration of this idea has been given in the 60s for composites of
free fields, namely to split a pointlike composite
\begin{equation}
:A^{2}(x):~\longrightarrow~A(x)A(y)
\end{equation}
by applying a subtle lightlike limiting procedure \cite{L-S} to the product
$:A^{2}(x)::A^{2}(y):$ which makes use of the singularity appearing for
lightlike separation. This idea was used in chiral models to show that
currents determine bilocals \cite{chiral}. But there is hardly any experience
with this splitting in gauge theories \cite{Jac}. A successful splitting would
of course automatically generate a gauge invariant bridged bilocal. It would
be a first step in an extension of the DHR superselection theory to gauge theories

The aim of this paper is to recall unexplored (and not explored) regions in
QFT (see its title) and shed new light onto them from the principle of
localization. All properties met in QFT models can an must be traced back to
this principle, only then one can claim to have understood the problem. It has
been shown elsewhere \cite{interface1} that QFT allows a presentation solely
in terms of the "modular positioning" of a finite number of "monads" where a
monad stands for the algebraic structure (hyperfinite type III$_{1}$ von
Neumann) which one meets in the form of localized algebras in QFT. The only
reason why this is mentioned here is its Leibnizian philosophical content: the
wealth of QFT can be encoded into the abstract positioning of a finite number
of copies of one monad into a common Hilbert space. The encoding encloses even
spacetime (the Poincare group representations) and the information about the
kind of quantum matter. In other words \textit{relative modular positions in
Hilbert space have physical reality, the substrate\footnote{Modular
positioning is the most radical form of relationalism since the local quantum
matter arises together with internal and spacetime symmetries. In other words
the concrete spacetime ordering is preempted in the abstract modular
positioning of the monads in the joint Hilbert space. } which is being
positioned does not.} Modular positioning, modular localization and
Poincar\'{e} symmetry are inexorably interwoven. This may sound provocative,
and certainly no practitioner would adopt or even sympatize with such an
extreme standpoint, but it is consistent with everything we know about QFT and
it constitutes the biggest difference to QM where none of this is realized. A
property encountered in a model of QFT has only been really understood, if it
has been traced back to the modular localization principle.

\section{Resum\'{e} and concluding remarks}

The main aim of this paper is the presentation of old important unsolved
problems of gauge theory in a new light. The standard gauge approach to the
renormalizable model of QED and its Yang-Mills generalization keeps the
pointlike formalism for s=1 vectorpotentials and sacrifices the Hilbert space
in intermediate computational steps whereas the new string-localized setting
avoids the indefinite metric Gupta-Bleuler/BRST formalism and all the
undesired aspects which come with it, as the fake pointlike localization of
gauge dependent operators and the impossibility to generate noncompact
localized physical operators from fake pointlike fields by implementing gauge
invariance (invariance under the BRST "symmetry"). In the case of the Higgs
mechanism there is in addition the necessity to pass between the unitary
(physical) and the renormalizable gauge. Another motivation comes from the
most attractive gauge of gauge theory namely the non-covariant axial gauge
which has the attractive property of coming with a Hilbert space
representation but has an incurable infrared divergency and for this reason
fell out of popularity with practitioners. The new viewpoint consists in
realizing that this gauge is not really a gauge in the standard use of the
terminology, rather it is a semiinfinite string-localized vectorpotential with
variable spacelike string directions which acts in a Hilbert space which
transforms covariant ($e$ plays an important role in the covariance law)

A more profound justification for the use of such comes from the fact that,
although certain covariant fields cannot exist in the setting of the Wigner
representation theory, the situation changes completely if one allows
semiinfinite spacelike string-localized covariant fields $\Psi(x,e)$ of scale
dimension 1, which we summarily called \textit{potentials}, since the
vectorpotential is the prime example (for higher $s$ there are also
tensorpotentials). These fields fulfill the correct power counting
prerequisite for renormalizability and do not need any power counting lowering
BRST formalism, not even in the massive case.

It may be helpful to collect the arguments for the use of those noncompact
localized potentials (instead of the pointlike indefinite metric potentials)
presented in this paper:

\begin{itemize}
\item The gauge theoretic argument why electrically charged operators cannot
be compactly localized remains obscure. Although the structural argument based
on Gauss's law is rigorous, it does not really explain the de-localization in
terms of localization aspects of the interaction.

\item Rewriting the quantum magnetic flux through a surface via Stokes theorem
into an integral over a pointlike vectorpotential leads to a contradiction
with the QFT A-B effect, whereas the use of string-localized vectorpotential
removes this discrepancy. Although this rather simple calculation does not
instruct how to formulate interactions, it does show that in order to avoid
incorrect conclusions about localizations, one must either return foe field
strengths or work with stringlike instead of pointlike vectorpotentials.

\item In most perturbative calculations in the gauge theoretical formalism the
condition of gauge invariance in terms of BRST invariance is clearly
formulated, but gauge invariant correlation functions of composite operators
(not to mention charged correlators) are, as a result of computational
difficulties, rarely calculated. In the new approach there is \textit{no gauge
conditions }to be imposed, rather the perturbative results are already the
physical one.

\item The reformulation of the Higgs phenomenon in the Schwinger-Higgs
screening setting removes some mysterious aspects of the former and brings it
into closer physical analogy with the Debeye screening mechanism of QM.
Whereas the latter explains how long range Coulomb interactions pass to
effective Yukawa potentials, the former describes the more radical change from
semiinfinite electrically charged strings with infrared photon clouds to
massive Wigner particles associated with pointlike fields. This more radical
screening is accompanied by a breaking of the charge symmetry (vanishing
charge) and the breaking of the even-oddness symmetry of the remaining real
field which makes the screening contribution from the alias charge neutral
$\varphi\varphi^{\ast}$ sector (after screening) indistinguishable from that
of $\varphi.$

\item The localization issue in case of Yang-Mills interactions and QCD (as
well as for selfinteracting $s\geq2$ models) is more involved since the change
under string direction is dynamical instead of the kinematical law
(\ref{change}) which follows from Wigner's representation theory. This leads
to a much stronger infrared behavior, in fact all spacetime correlators are
infrared divergent and only some spacetime independent coupling-dependent
functions as the beta function are infrared finite. The new
string-localization approach explains this and proposes to take care of the
$e$-fluctuations which cause the infrared divergences and clarify their role
in confinement/invisibility and gauge-bridge breaking (jet formation). This,
as well as the still missing presentation of an Epstein-Glaser approach in the
presence of string potentials, will be the topic of a separate
work\footnote{Jens Mund and Bert Schroer, in progress.}.

\item The approach based on string-localized potentials does not only replace
the gauge setting, which resulted from a resolution of the clash between
pointlike localization and quantum positivity with the brute force method of
indefinite metric, but it is also meant to be useful for higher spins
(example: $g_{\mu\nu}$ string tensorpotentials) where such a gauge trick is
not known. In addition to the avoidance of indefinite metric it also lowers
the short distance dimension of pointlike field strength s+1 (for spin s) to
the lowest value $d_{sca}=1$ allowed by unitarity which is the prerequisite of
having renormalizable interactions for any spin.
\end{itemize}

Within the conventional standard terminology of QFT the present project to
incorporate string-localized objects into already existing settings (standard
model, s=2 "gravitons") would be considered as "nonlocal" QFT. To make QFT
compatible with nonlocality is one of the oldest projects of relativistic QT.
Apart from early (pre-renormalization) attempts to modify the quantum
mechanical commutation relations to make them more quantum-gravity friendly,
the more systematic investigations in QFT started in the 50s with attempts by
Christensen and M\o ller to improve the behavior of interactions in the
ultraviolet by spreading interaction vertices in a covariant manner. Later
attempts included the Lee-Wick proposal to modify Feynman rules by pair of
complex poles and their conjugates. All these models were eventually shown to
contradict basic macro-causality properties which are indispensable for their
interpretation\cite{Anth}. There are of course relativistic quantum mechanical
theories which lead to a Poincar\'{e}-invariant clustering S-matrix
\cite{interface1}, but do not fit into the causal localization of the QFT setting.

The more recent interest in nonlocal aspects originated in ideas about
algebraic structures (noncommutative QFT) which are supposed to replace
classical spacetime as the first step towards "quantum gravity". These
attempts usually start from a modification of the quantum mechanical
uncertainty relation which, since they involve position operators, strictly
speaking does not exist in QFT\footnote{As a result these attempts lead to
problems with the principle of independence of the reference frame and in a
certain sense open the backdoor for the return of the ether.}. The only analog
of the uncertainty relation which comes to one's mind is the statement that
one can associate to a localized algebra $\mathcal{A(O})$ and a "collar" of
size $\varepsilon$ (the splitting distance) which separates $\mathcal{A(O})$
from its causal disjoint \cite{Haag}, a localization entropy (or energy)
$Ent(\varepsilon)$ which is proportional to the surface and diverges for
$\varepsilon\rightarrow0$ in a model-independent manner \cite{BMS}. Whether
such relations between the sharpness of localization and the increase of
entropy/energy can be the start of a noncommutative/nonlocal project remains
to be seen.

If one can consider this thermal relation as a QFT substitute of an
uncertainty relation, it points into a quite different direction than the
proposal for a more noncommutative modification of QFT \cite{Bahns}. Namely it
looks like an invitation to explore connections between
thermodynamics/statistical mechanics, a project which Ted Jacobson has pursued
for some time \cite{Ted}. Unlike algebraic modifications for position
operators it has the appealing feature of not having to struggle with problems
of frame dependence.

The philosophy underlying the noncommutative approach has been nicely exposed
in a recent essay by Sergio Doplicher \cite{Do}. His emphasis that a principle
as causal localization can only be overcome by another principle which
contains the known one in the limiting situation of large distances is
certainly well taken, as in many cases, the devil is in the details.

In the present work, the nonlocal behavior remains part of QFT; it may go
against certain formalisms as Lagrangian quantization or functional
representations, but it certainly does not lead to reasons to change the
principles of QFT and it also is not "revolutionary", it only belongs to one
of its unexplored corners. Even with respect to quantum gravity the two
nonlocal approaches remain different. Within the nonlocality allowed by QFT it
would be tempting to relate gravity with selfinteracting $d_{sca}=1$
string-localized tensor potentials $g_{\mu\nu}(x,e).$ The hope is that one can
access this problem (and if necessary dismiss it) with more conventional means.

The aim of this paper was to cast new light on unexplored regions of gauge
theory based on recent progress in the understanding of modular localization.
There was however no attempt to go into the important details of the new
perturbation theory in terms of string-localized potentials. This will be the
subject of forthcoming work \cite{Jens}.

\textit{Acknowledgements}: I thank Jakob Yngvason for his invitation to visit
the ESI in Vienna and for his hospitality. Part of the research which entered
this paper was carried out at the ESI. I also acknowledge several discussions
with Jakob Yngvason especially on matters related to section 5.

This work is in a way a continuation of a paper written several years ago
together with Jens Mund and Jakob Yngvason. I am indebted to Jens Mund with
whom I have maintained close scientific contact and who has kept me informed
about recent results about attempts to adapt the Epstein-Glaser
renormalization and the idea of wave front sets to string-localized fields. I
owe Karl-Henning Rehren thanks for bringing me up to date on matters which
were relevant for the appendix.

\section{Appendix}

\subsubsection{Modular group for conformal algebras localized in
doubly-connected spacetime rings}

It has been known for a long time that the modular group for a conformal
double cone which is placed symmetrically around the origin is related to that
of its two-dimensional counterpart by rotational symmetry. In other words if%
\begin{equation}
x_{\pm}(\tau)=\frac{(1+x_{\pm})-e^{\tau}(1-x_{\pm})}{(1+x_{\pm})+e^{\tau
}(1-x_{\pm})},~-1<x_{\pm}<1
\end{equation}
represents the two-dimensional conformal modular group in lightray coordinates
for a two-dimensional double cone symmetrically around the origin, then the
modular group of a symmetrically placed four-dimensional double cone which
results from the two-dimensional region by rotational symmetry acts as above
by simply replacing $x_{\pm}(\tau)$ in the above formula by their radial
counterpart \cite{H-L}
\begin{equation}
x_{\pm}(\tau)\rightarrow r_{\pm}(\tau),\text{ }\varphi,\theta,~\tau
-independent \label{rad}%
\end{equation}

The generalization to two and more copies of double cones in two dimensions,
symmetrically placed on both sides of the origin is obviously a group which in
terms of $x_{\pm}$ has 4 or 2n fixed points which are the endpoints of two
separated intervals. The construction of explicit formulae for n intervals
$E=$ $I_{1}\cup I_{2}..\cup I_{n}~$with 2n fixed points is well-known; they
are most conveniently obtained as Caley-transforms of one-parametric subgroups
of $Diff(S^{1})$%
\begin{align}
&  f_{\tau}^{(n)}(z)=\sqrt[n]{Dil(-2\pi\tau)z^{n}},~~Dil(-2\pi t)x=e^{2\pi
t}x\\
&  x\rightarrow z=\frac{1+ix}{1-ix},\text{ }Cayley~transf.~\dot{R}\rightarrow
S^{1}\nonumber
\end{align}
This diffeomorphism group in terms of x is infinity-preserving. By applying
further infinity preserving symmetry transformations (translations, dilations)
we may achieve the desired symmetric situation with respect to the origin.

For n=2 the two double cones are the x-t projections of 4-dimensional matter
localized in $\mathcal{T}$ $\ $and not matter in a 2-dimensional conformal
theory. This suggests that in looking for a geometric analog of (\ref{rad})
one should be aware that the full diffeomorphism group $Diff(S^{1})$ has no
analog in 4 dimensions; in fact not even the Moebius subgroup associated to
the Virasoro generator $L_{0}$ has a counterpart. Hence arguments which are
based on properties of $L_{0}$ as the necessity to work with split vacua
states \cite{KL} or with "mixing" \cite{Ca-Hu}\cite{LMR} are not applicable here.

The use of the above formalism in connection with modular theory of
multi-intervals\ and two-dimensional multi-double cones has been presented in
detail in \cite{KL}. In particular it was shown that in the presence of the
$L_{0}$ in the Virasoro algebra there is no global representation of the
$f_{\tau}^{(n)}(z)$ diffeomorphism. Rather the best one can do by choosing
instead of the global vacuum the so-called split vacuum is to represent this
diffeomophism group on $E$ and have a non-geometric action on its complement
$E^{\prime},$ or construct a "geometric state" (another split vacuum) for
$E^{\prime}$ and find a nongeometric action on $E.$

In the special case of a chiral Fermion one can achieve a global
quasi-geometric action in the vacuum at the expense of a mixing between the
different intervals by a computable mixing matrix \cite{Ca-Hu}\cite{LMR}. But
only the projections of localized zero mass matter in d=1+3 are candidates for
a pure geometric action in their standard vacuum state.

This difference extends to the explanation of violation of Haag duality for
$(m=0,s\geq1).$ Whereas in the chiral case this is due to charge transporters
whose construction requires the setting of field theory with its
characteristic property of vacuum polarization, the Aharonov Bohm effect in
QFT (and its higher spin s%
$>$%
1 generalization) can be fully described in the Wigner one-particle
representation. It is the only known topological effect in QFT which is of a
porely classical origin.

The localization of n $\mathcal{T}$ symmetrically placed around the origin has
a x-t projection which consists of n symmetrically arranged two-dimensional
double cones. The diffeomorphism group which leaves this figure invariant is a
particular diffeomorphism group which in lightray coordinates is a
diffeomorphism with 2n fixed points. The number of stringlike potential
associated with a pointlike field strength increases with spin s; there is
always one with the lowest possible dimension which is $d_{sca}$=1 and the one
with the highest dimension has a $d_{sca}$ which is smaller than that of the
lowest field strength. So the A-B fluxes which account for the
string-localized potentials are certainly expected to increase with s. But the
situation of n disconnected $\mathcal{T}$ appears repetitive. It would be
fascinating if the increase of s could be linked with the occurrence of a new
type of A-B effect in higher genus (higher connectivity) analogs of
$\mathcal{T}$ instead of being n-$\mathcal{T}$ repetitive.

\ 
\end{document}